\documentclass[12pt,a4paper]{article}
\usepackage[a4paper]{geometry}
\usepackage[english]{babel}
\usepackage[cp850]{inputenc}
\usepackage[dvips]{graphicx}
\usepackage{amsmath,amsfonts,amsthm,amssymb}
\usepackage{mathrsfs}
\usepackage[usenames, dvipsnames]{color}
\usepackage[colorlinks=true,citecolor=blue,linkcolor=blue,urlcolor=blue,pdfstartview=FitH]{hyperref}
\usepackage{nameref}
\usepackage[svgnames]{xcolor}

\usepackage{geometry}
\usepackage{tikz,tkz-tab}
\usetikzlibrary{fit,calc,positioning}

\graphicspath{{images/}}

\DeclareGraphicsExtensions{.pdf,.png}

\makeatletter
\newcommand{\manuallabel}[2]{\def\@currentlabel{#2}\label{#1}}
\makeatother

\newtheorem{framework}{Framework}

\def\Imb{\mbox{Imb}}
\def\Esp{\mathbb{E}}
\def\calT{{\cal T}}
\def\cardT{{{\rm Card}(\calT}}
\def\é{\'{e}}
\def\è{\`{e}}
\def\ê{\^{e}}
\def\à{\`{a}}
\def\ô{\^{o}}

\title{Limit Order Strategic Placement\\ with Adverse Selection Risk\\ and the Role of Latency.}
\author{Charles-Albert Lehalle\thanks{Capital Fund Management, Paris and Imperial College, London}{} ~%
  and %
  Othmane Mounjid\thanks{Universit\'e Pierre et Marie Curie, Paris}}
\date{Printed the \today}
\begin{document}
\maketitle

\begin{abstract}
  This paper is split in three parts: first, we use labelled trade data to exhibit how market participants decisions depend on liquidity imbalance; then, we develop a stochastic control framework where agents monitor limit orders, by exploiting liquidity imbalance, to reduce adverse selection. For limit orders, we need optimal strategies essentially to find a balance between fast execution and avoiding adverse selection: if the price has chances to go down the probability to be filled is high but it is better to wait a little more to get a better price. In a third part, we show how the added value of exploiting liquidity imbalance is eroded by latency: being able to predict future liquidity consuming flows is of less use if you do not have enough time to cancel and reinsert your limit orders. There is thus a rationale for market makers to be as fast as possible to reduce adverse selection.
Latency costs of our limit order driven strategy can be measured numerically.

To authors' knowledge this paper is the first to make the connection between empirical evidences, a stochastic framework for limit orders including adverse selection, and the cost of latency. Our work is a first step to shed light on the role played by latency and adverse selection in optimal limit order placement.
\end{abstract}

\section{Introduction}
\label{sec:1}
With the electronification, fragmentation, and increase of trading frequency, orderbook dynamics is under scrutiny. Indeed, a deep understanding of orderbook dynamics provides insights on the price formation process. There is essentially two approaches for modelling the price formation process. First, \emph{general equilibrium models} based on interactions between rational agents who take optimal decisions. General equilibrium models focus on agents behaviours and interactions. For example, investors split their metaorders into large collections of limit orders (i.e. liquidity providing) and market orders (i.e. liquidity consuming) (see \cite{citeulike:3320208}, \cite{citeulike:13177976}, \cite{citeulike:13497373}) while (high frequency) market makers mostly use limit orders to provide liquidity to child orders of investors (see \cite{citeulike:13675263,citeulike:13497022}).  Second, \emph{statistical models} where the orderbook is seen as a random process (see \cite{citeulike:6032638}, \cite{citeulike:6659908}, \cite{citeulike:12810809} and references herein). Statistical models focus on reproducing many salient features of real markets rather than agents' behaviours and interactions. In this paper, we consider a statistical model where the arrival and cancellation flows follow size dependent Poisson processes. Using this model, we setup an optimal control for one agent targeting to obtain the ``best price'' by maintaining optimally a limit order in the orderbook .\\

In practice, market participants use optimal trading strategies to find a balance between at least three factors: the price variation uncertainty, the market impact and the inventory risk. For example, an asset manager who took the decision to buy or sell a large number of shares needs to adapt its execution speed to price variations. The simplest case would be to accelerate execution when price moves in its favor. He needs  to consider the market impact too and in particular the price pressure of large orders: fast execution of huge quantities consumes orderbook liquidity and increases transaction costs. Finally, there is an inventory risk associated to the orders size: it is riskier to hold a large position than a small one during the same period of time. A fast execution reduces this inventory risk. The asset manager should then find the optimal balance between trading slow and fast. Models for these strategies are now well known (see \cite{citeulike:9304794} and \cite{GLFT} ).
Recent papers introduce a risk term in their optimisation problem. Moreover, some papers combine even short term anticipations of price dynamics inside these risk control frameworks. For example, in \cite{ALOR06}, authors include a Bayesian estimator of the price trend in a mean-variance optimal trading strategy. In \cite{citeulike:13587586}, authors include an estimate of futur liquidity consumption -- $\mu$ in their paper shoud be compared to our consuming intensities $\lambda^{\cdot,-}$ -- in macroscopic optimal execution.

In this paper, we consider an optimal control problem where the agent faces the price variation uncertainty and the market impact but there is no inventory risk since we consider one limit order. The idea is to propose optimal strategies that can be plugged into any large scale strategy (see \cite[Chapter 3]{citeulike:12047995} for a practitioner viewpoint on splitting the two time scales of metaorders executions) by taking profit of a short term anticipation of price moves.

After some considerations about short time price predictions and empirical evidences showing that market participants decisions depend on the imbalance (see Section \ref{sec:empirics}), we show that optimal control can add value to any short term predictor (see Section \ref{sec:understanding}) in the context of simple control (cancel or insert a limit order) for a ``large tick'' stock.
This result can thus be used by investors or market makers to include some predictive power in their optimal trading strategies.

Then we show how latency influences the efficient use of such predictions. Indeed, the added value of the optimal control is eroded by latency. It allows us to link our work to regulatory questions. First of all: what is the ``value'' of latency? Regulators could hence rely on our results to take decisions about ``slowing down'' or not the market (see \cite{citeulike:13586167} and \cite{citeulike:12721030} for discussions about this topic).
It sheds also light on maker-taker fees since the real value of limit orders (including adverse selection costs), are of importance in this debate (see \cite{harris2013maker} for a discussion).

This paper can be seen as a mix of two early works presented at the ``Market Microstructure: Confronting Many Viewpoints'' conference (Paris, 2014):
a data-driven one focused on the predictive power of orderbooks \cite{sasha14imb}, and an optimal control driven one \cite{moa14hjb}.
Our added values are first a proper combination of the two aspects (inclusion of an imbalance signal in an optimal control framework for limit orders), and then the construction of our cost function. Unlike in the second work, we do not value a transaction with respect to the mid-price at $t=0$, but with respect to the microprice (i.e. the expected future price given the liquidity imbalance) at $t=+\infty$. We will argue the difference is of paramount importance since it introduces an effect close to adverse selection aversion, that is crucial in practice.

As an introduction of our framework, we will use a database of labelled transactions on NASDAQ OMX (the main Nordic European regulated markets) to show how orderbook imbalance is used by market participants in a way that can be seen as compatible with our theoretical results.

Hence the structure of this paper is as follows:
Section \ref{sec:empirics} presents orderbook imbalance as a microprice and illustrates the use of imbalance by market participants thanks to the NASDAQ OMX database.
Once these elements are in place, Section \ref{sec:dpp} presents our model and \ref{sec:understanding} shows how to numerically solve the control problem and provides main results, especially the influence of latency on the efficiency of the strategy.


%

%
\section{Main Hypothesis and Empirical Evidences}
\label{sec:empirics}

\subsection{Database presentation}
The data used here are a direct feed on NASDAQ-OMX, that is the primary market\footnote{i.e. the \emph{regulated exchange} in the MiFID sense, see \cite{citeulike:12047995} for details.} on the considered stock. 
Capital Fund Management feed recordings for AstraZeneca accounts for 72\% of market share (in traded value) for the continuous auction on this stock over the considered period.
Surprisingly, there is currently no academic paper comparing the predictive power of imbalances of different trading venues on the same stock. It is outside of the scope of this paper to elaborate on this. We will hence consider the liquidity on our primary market is representative of the state of the liquidity on other ``large'' venues (namely Chi-X, BATS and Turquoise on the considered stock). If it is not the case it will nevertheless not be difficult to adapt our result relying on statistics on each venue, or on the aggregation of all venues.
We did not aggregated venues ourselves for obvious synchronization reasons: we do not know the capability of each market participant to synchronize information coming from all venues and do not want to add noise by making more assumptions. Our idea here is to use the state of liquidity at the first limits on the primary market as a proxy of information about liquidity really used by participants .\\
\medskip
\begin{table}[!ht]
  \centering
  \begin{tabular}{|l|r|r|}\hline
    Venue & AstraZeneca & Vodafone \\ \hline
    BATS Europe   &  7.16\% & 7.63\%\\
    Chi-X               & 19.27\% & 20.02\%\\
    Primary market          &  72.24\% & 61.09\%\\
    Turquoise    &   1.33\% & 11.26\% \\\hline
  \end{tabular}
  \caption{Fragmentation of AstraZeneca (compared to Vodafone) from the 2013-01-02 to the 2013-09-30}
  \label{tab:frag}
\end{table}
We focus on NASDAQ-OMX because this European market has an interesting property: market members' identity is known. It implies transactions are labeled by the buyer's and the seller's names. Almost all trading on NASDAQ Nordic stocks was labelled this way until end of 2014 (more details are available in \cite{citeulike:13497022}, because this whole paper is based on this labelling).
Note members' identity is not investors' names; it is the identity of brokers or market participants large enough to apply for a membership. High Frequency Participants (HFP) are of this kind. Of course some participants (like large asset management institutions) use multiple brokers, or a combination of brokers and their own membership.
Nevertheless, one can expect to observe different behaviours when members are different enough. We will here focus on three classes of participants: High Frequency Participants (HFP), global investment banks, and regional investment banks.

\subsection{Stylized fact 1:  the predictive power of the orderbook imbalance}

\paragraph{Short term price prediction utility.} Academic papers (see \cite{citeulike:6659908} or \cite{citeulike:12810809}) and brokers' research papers (see \cite{bes16imb}) document how the sizes at first limits of the public orderbook\footnote{Limit orderbooks are used in electronic market to store unmatched liquidity, the \emph{bid size} is the one of passive buyers and the \emph{ask size} the one of passive sellers; see \cite{citeulike:12047995} for detailed.} influence the next price move.
It is worthwhile to underline that the identified effects are usually not strong enough to be the source of a statistical arbitrage: the expected value of buying and selling back using accurate predictions based on sizes at first limits does not beat transaction costs (bid-ask spread and fees). See \cite{citeulike:13027037} for a discussion.
Nevertheless:
\begin{itemize}
\item For an investor who \emph{already took the decision to buy or to sell}, this information can spare some basis points. For very large orders, it makes a lot of money and in any case it reduces implicit transaction costs.
\item Market makers naturally use this kind of information to add value to their trading processes (see \cite{citeulike:12335801} for a model supporting a theoretical optimal market making framework including first limit prices dynamics).
\end{itemize}

The easiest way to summarize the state of the orderbook without destroying its informational content is to compute its \emph{imbalance}: the quantity at the best bid minus the one at the best ask divided by the sum of these two quantities:
\begin{equation}
  \label{eq:imbdef}
  \Imb_t := \frac{Q^{Bid}_t - Q^{Ask}_t}{Q^{Bid}_t + Q^{Ask}_t}.
\end{equation}

\begin{figure}[!ht]
  \centering
  \hfill (a)\hfill (b)\hfill~\\
    \includegraphics[width=.45\linewidth]{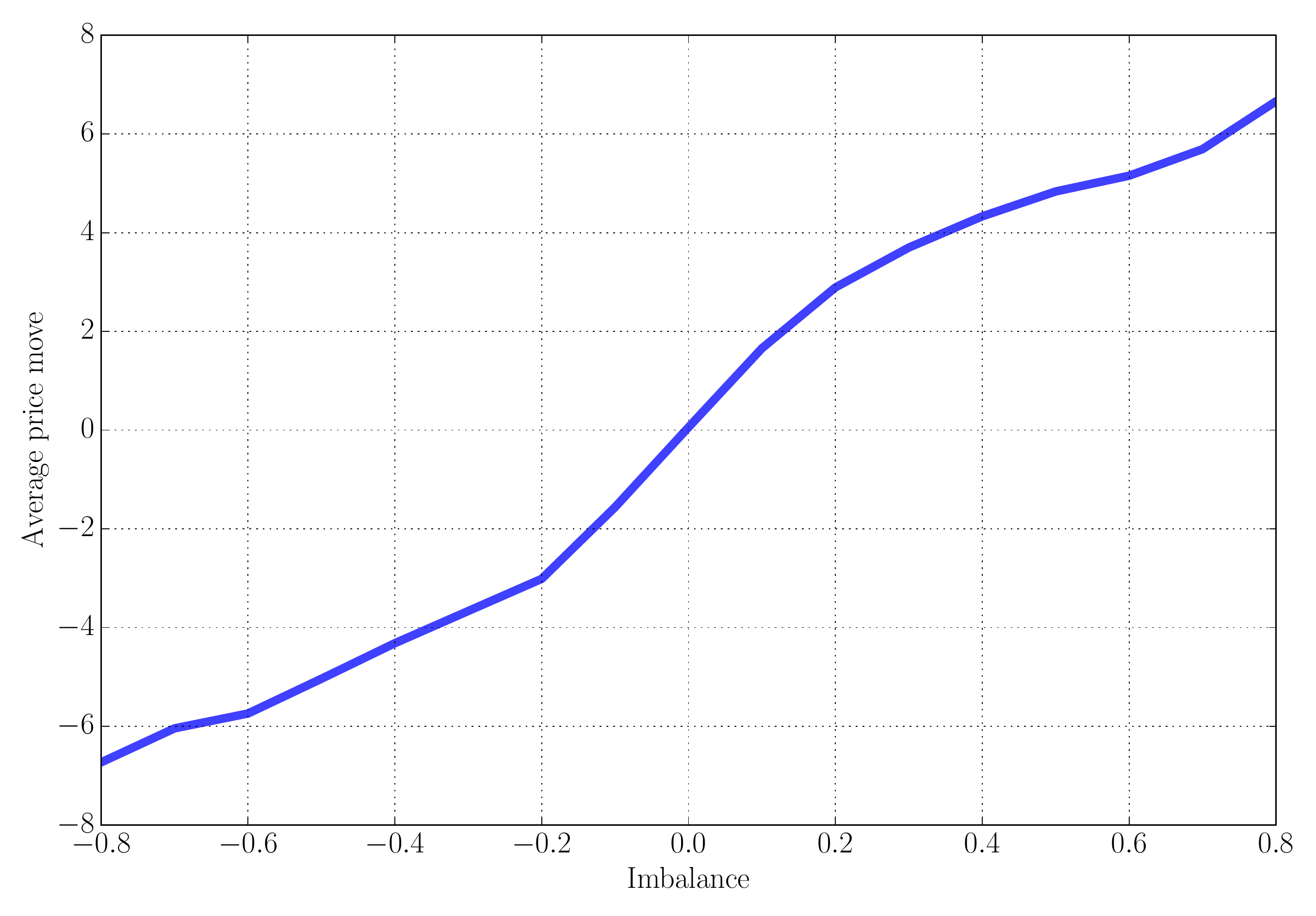}\hfill
  \includegraphics[width=.45\linewidth]{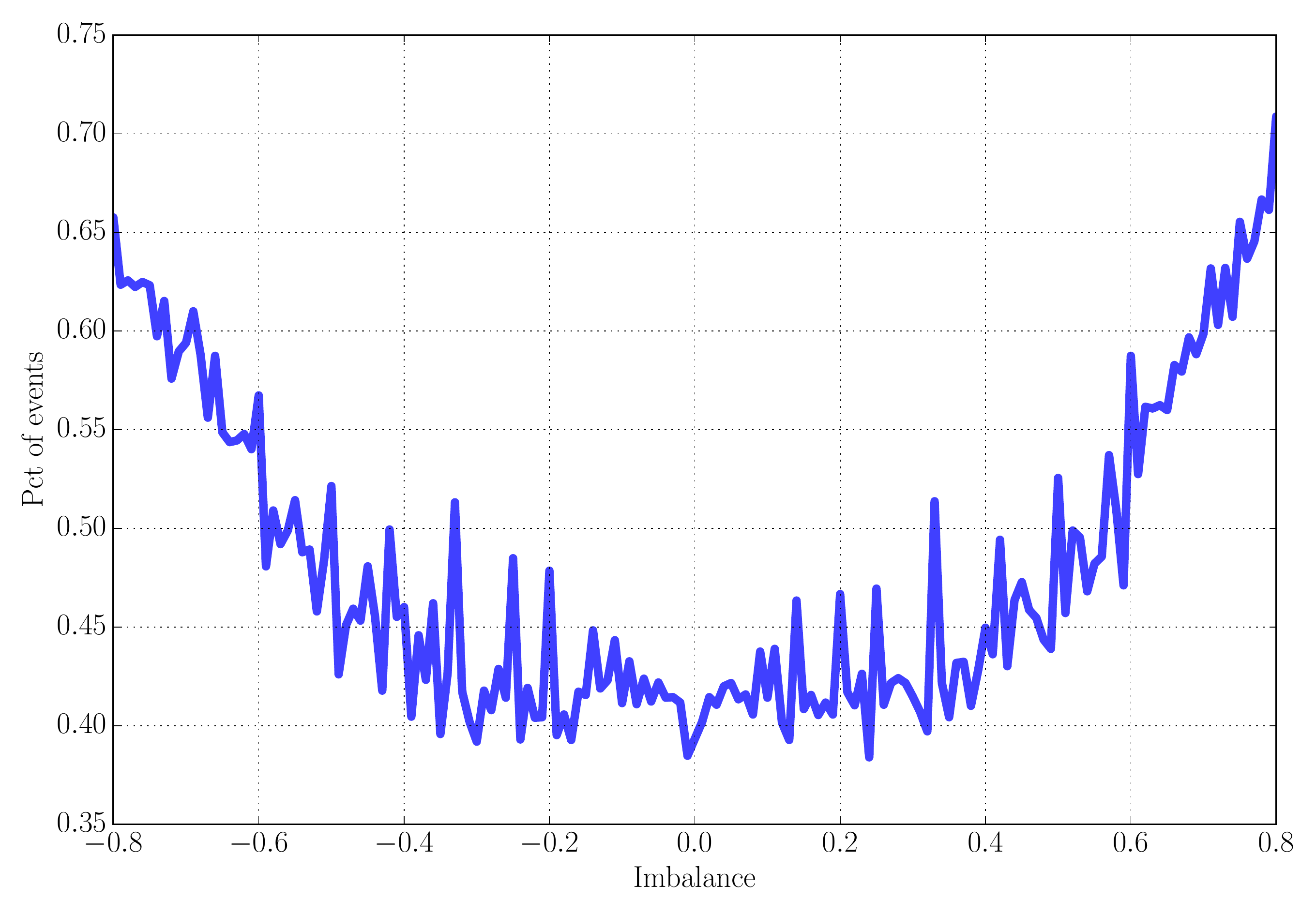}\\
  \caption{(a) The predictive power of imbalance on stock Astra Zeneca: imbalance (just before a trade) on the $x$-axis and the expected price move (during the next 50 trades) on the $y$-axis. (b) distribution of the imbalance just before a trade. From the 2013-01-02 to the 2013-09-30 (accounts for 376,672 trades).}
  \label{fig:predpowimb}
\end{figure}

\paragraph{The nature of the predictive power of the imbalance.}
The predictive power of the orderbook imbalance is well known (see \cite{citeulike:12820703}).
The rationale of this stylized fact (i.e. \emph{the midprice will go in the direction of the smaller size of the orderbook}) is outside of the scope of this paper. We just give here some clues and intuitions to the reader:
\begin{itemize}
\item The future price move is positively correlated with the imbalance.
In other terms
\begin{equation*}
  \label{eq:imbpred}
  \Esp( (P_{t+\delta t} - P_t) \times \mbox{sign}(\Imb_t) | \Imb_t) > 0,
\end{equation*}
where $P_t$ is the midprice (i.e. $P_t = (P^{Bid}_t + P_t^{Ask})/2$, where $P^{Bid}$ and $P^{Ask}$ are respectively the best bid ans ask prices) at $t$ for any $\delta t$. Obviously when $\delta t$ is very large, this expected price move is very difficult to distinguish from large scale sources of uncertainty.
See for instance \cite{citeulike:12820703} for details on the ``predictive power'' of such an indicator
(our Figure \ref{fig:predpowimb} illustrates this predictive power on real data).
\item Within a model in which the arrival and cancellation flows  follow independent point processes of the same intensity, the smallest queue (bid or ask) will be consumed first, and the price will be pushed in its direction. 
  See \cite{citeulike:12810809} for a more sophisticated point process-driven model and associated empirical evidences.
\item Another viewpoint on imbalance would be that the bid vs. ask imbalance contains information about the direction of the net value of investors' metaorders : first, in a direct way if one is convinced that investors post limit orders; second, indirectly if one believes investors only consume liquidity and in such a case bid and ask sizes are an indicators of market makers net inventory.
\end{itemize}
The focus of formula \eqref{eq:imbdef} on two first limits weaken the predictive power of the bid vs. ask imbalance. For large tick assets\footnote{For a focus on tick size, see \cite{doi:10.1142/S2382626615500033}.} it may be enough to just use the first limits, but for small tick ones it certainly increases the predictive power of our imbalance indicator to take more than one tick into account.
Since a discussion on the predictive power of imbalance is outside the scope of the paper, we will stop here the discussion.
\medskip

\paragraph{Empirical evidences on the predictive power of the imbalance.}
Figure \ref{fig:predpowimb}.a shows the imbalance \eqref{eq:imbdef} on the $x$-axis and the midprice move 
after 50 trades on the $y$-axis. In Figure \ref{fig:predpowimb}.a, we recover that the imbalance is highly linearly positively correlated to the price move after 50 trades. Figure \ref{fig:predpowimb}.b shows the distribution of imbalance just before a change in the orderbook state.  In Figure \ref{fig:predpowimb}.b, agents are highly active at extreme imbalance values. People become highly active at extreme imbalances because they identify a profit opportunity to catch or at the opposite an adverse selection effect to avoid.  Another explanation may come from  the predictive power of the imbalance (see Figure \ref{fig:predpowimb}.a). In fact, participants start to anticipate the next price move when the signal imbalance is strong while they are inactive when they have no view on the next price move (i.e the signal imbalance is weak). 

\medskip

This paper provides a stochastic control framework to post limit orders using the information contained in the orderbook imbalance.
In such a context, we will call the \emph{microprice seen from $t$} and note $P_{+\infty}(t)$:
\begin{equation*}
  \label{eq:micropricedef}
  P_{+\infty}(t) =  \lim_{\delta t\rightarrow+\infty}\Esp( P_{t+\delta t}  | P_t, \Imb_t).
\end{equation*}

\subsection{Stylized fact 2: Agent's decisions depend on the orderbook liquidity} 
\subsubsection{Agent's decisions depend on the orderbook imbalance}
We expect some market participants to invest in access to data and technology to take profit of the informational content of the orderbook imbalance. 
A very simple way to test this hypothesis is to look at the orderbook imbalance just before a transaction with a limit order for a given class of participant. We will focus on three classes of \emph{agents} (i.e. market participants): Global Investment Banks, High Frequency Participants (HFP), and Regional Investment Banks or Brokers.
Table \ref{tab:statdesa} provides descriptive statistics on these classes of participants in the considered database.

\begin{table}[!ht]
  \centering
  \begin{tabular}{|llr|rr|}\hline
Order type & Participant type & Order side & Avg. Imbalance & Nbe of events \\\hline\hline
  Limit   &   Global Banks  &    Sell     &       0.35  & 62,111\\ 
               &             &  Buy         &   -0.38 &  63,566\\\hline 
        &   HFP           &    Sell     &       0.32  & 52,315\\ 
        &                   &   Buy       &     -0.33 &  46,875\\\hline 
         &  Instit. Brokers  & Sell    &        0.57  &  6,226\\ 
          &                   & Buy        &    -0.52   & 4,646\\\hline 
  \end{tabular}
  \caption{Descriptive statistics for our three classes of agent. AstraZeneca (2013-01 to 2013-09).}
  \label{tab:statdesa}
\end{table}

We focus on limit orders since information processing, strategy and latency play a more important role for such orders than for market orders (market orders can be sent blindly, just to finish a small metaorder or to cope with metaorders late on schedule, see \cite{citeulike:12047995} for ellaborations on brokers' trading strategies).
\medskip

For the following charts, we use labelled transactions from NASDAQ-OMX\footnote{For each transaction, we have a buyer ID, seller ID, a size, a price and a timestamp.} and thanks to timestamps (and matching of prices and quantities) we synchronize them with orderbook data (recorded from direct feeds by Capital Fund Management). It enables us to snapshot the sizes at first limits on NASDAQ-OMX just before the transaction.

Say for a given participant (i.e. \emph{agent}) $a$ the quantity at the best bid (respectively best ask) is $Q^{Bid}_\tau(a)$ (resp. $Q^{Ask}_\tau(a)$) just before a transaction at time $\tau$ involving a limit order owned by $a$. 
We note $Q^{same}_\tau(a):=Q^{Bid}_\tau(a)$ (respectively $Q^{opposite}_\tau(a):=Q^{Ask}_\tau(a)$) for a buy limit order and $Q^{same}_\tau(a):=Q^{Ask}_\tau(a)$ (respectively $Q^{opposite}_\tau(a):=Q^{Bid}_\tau(a)$) for a sell limit order. 

We normalize the quantities by the best opposite to obtain $\rho_\tau(a)=\cfrac{Q^{same}_\tau(a)-Q^{opposite}_\tau(a)}{Q^{same}_\tau(a)+Q^{opposite}_\tau(a)}$. It is then easy to average over the transactions indexed by timestamps $\tau$ to obtain an estimate of this expected ratio for one class of agent:
\begin{equation*}
  \label{eq:rhoa}
  R(a)=\frac{1}{\cardT)} \sum_{\tau\in\calT} \rho_\tau(a),\; \lim_{\cardT)\rightarrow+\infty} R(a) = \Esp_\tau \left( \frac{Q^{same}_\tau(a)-Q^{opposite}_\tau(a)}{Q^{opposite}_\tau(a)+Q^{opposite}_\tau(a)} \right).
\end{equation*}
It is even possible to control a potential bias by using the same number of buy and sell executed limit orders to compute this ``neutralized'' average:
\begin{equation*}
  \label{eq:rhoa}
  R'(a)=\frac{1}{\cardT({\rm buy}))} \sum_{\tau\in\calT({\rm buy})} \rho_\tau(a) + \frac{1}{\cardT({\rm sell}))} \sum_{\tau\in\calT({\rm sell})} \rho_\tau(a). %
\end{equation*}

\begin{figure}[!ht]
  \centering
  \hfill (a)\hfill \hfill (b)\hfill~\\
  \includegraphics[width=.45\linewidth]{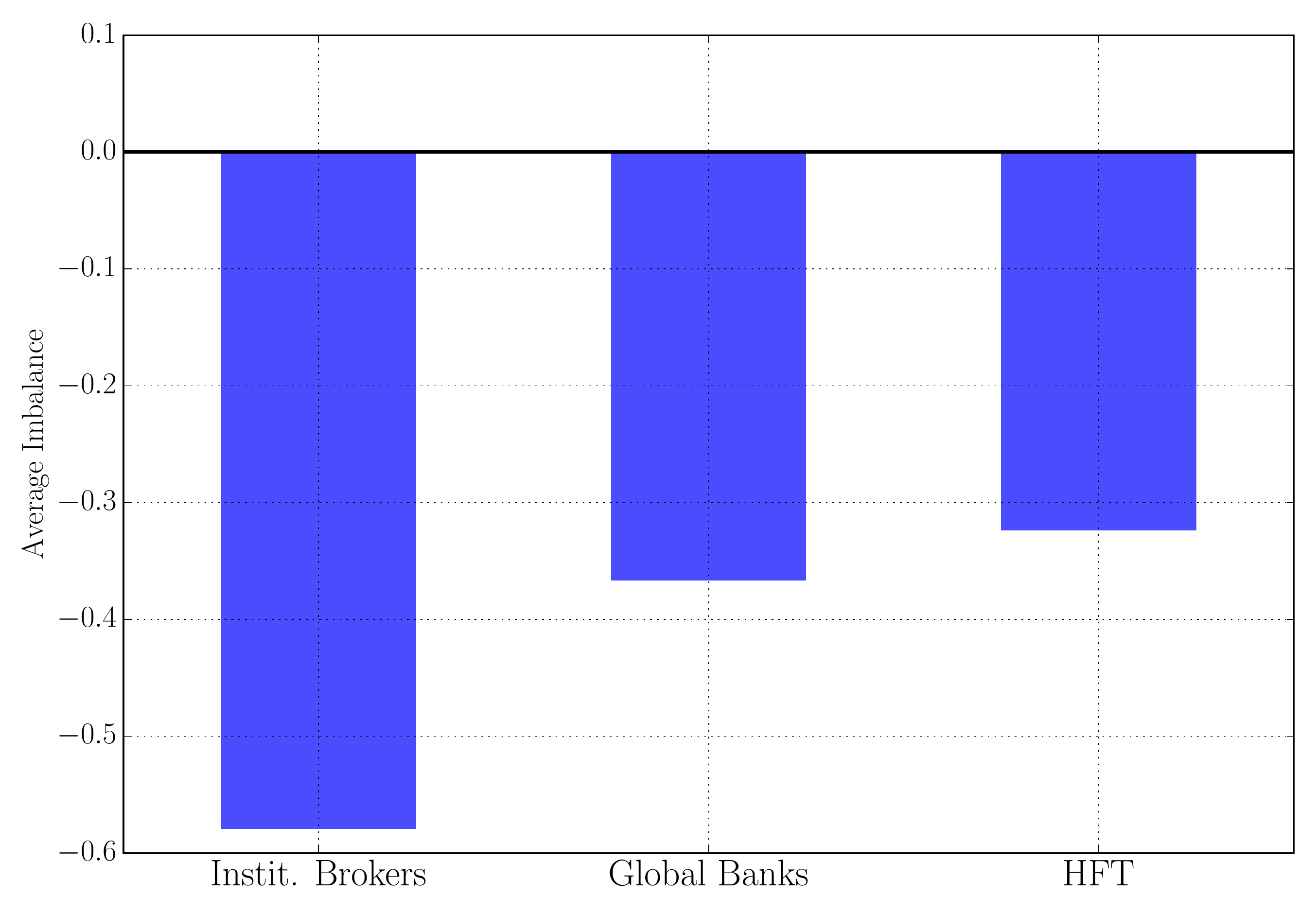}\hfill
  \includegraphics[width=.45\linewidth]{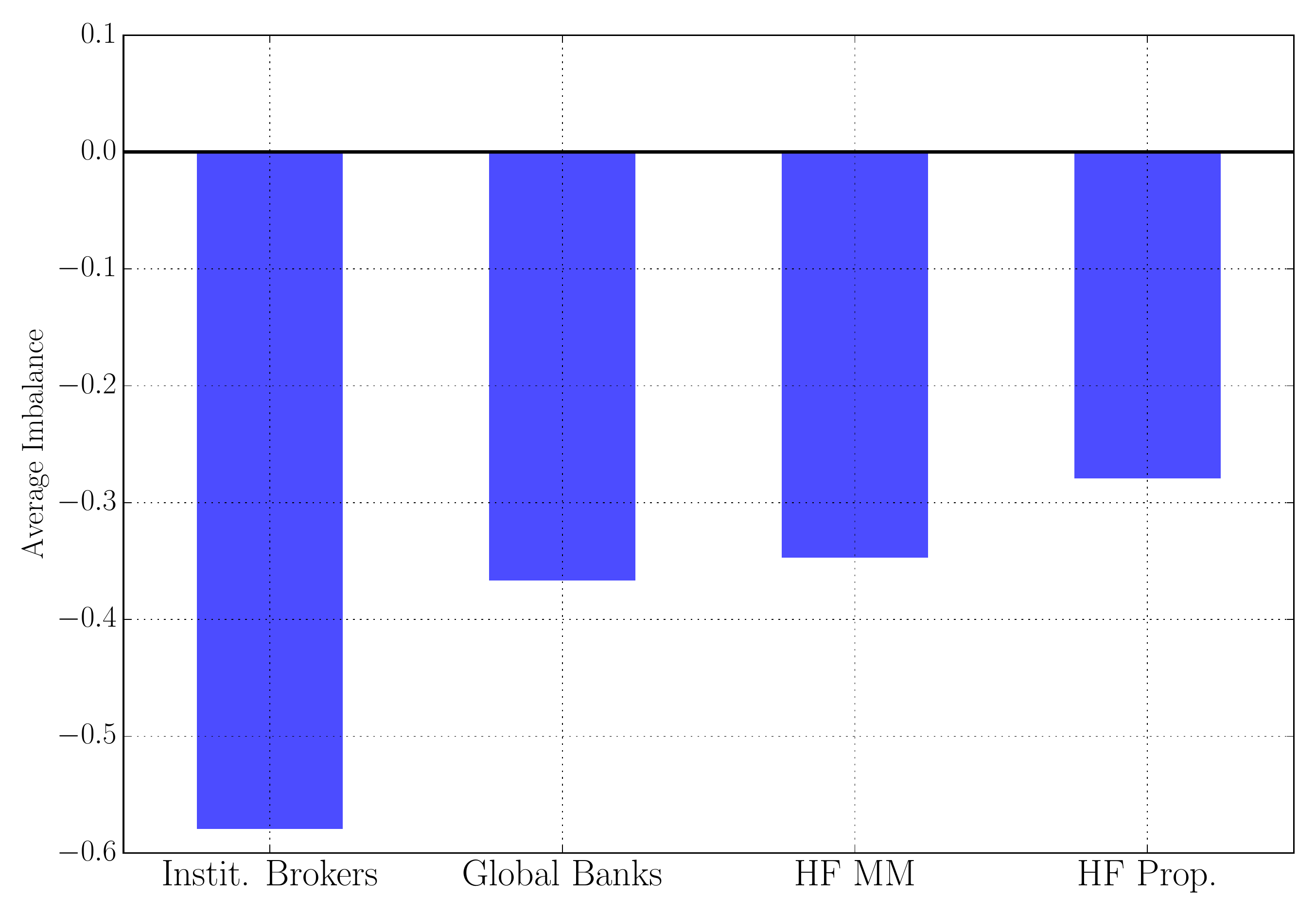}\\
  \caption{Comparison of neutralized orderbook Imbalance $R'(a)$ at the time of a trade via a limit order (a) for institutional brokers, global investment banks and High Frequency Participants. (b) Shows a split of HFP between market makers and proprietary traders. Data are the ones for AstraZeneca (2013-01 to 2013-09).}
  \label{fig:imbNASDAQ}
\end{figure}

Figure \ref{fig:imbNASDAQ}.a shows the average state of the imbalance (via some estimates of $R'(a)$, on AstraZeneca from January 2013 to August 2013) for each class of agent (see Tables \ref{tab:compagents:HFT}, \ref{tab:compagents:GIB}, \ref{tab:compagents:brok} for lists of NASDAQ-OMX memberships used to identify agents classes).
One can see the state of the imbalance is different for each class given it ``accepted'' to transact via a limit order:
\begin{itemize}
\item Institutional brokers accept a transaction when the imbalance is largely negative, i.e. they buy using a limit order while the price is going down. It generates a large adverse selection: they would have wait a little more, the price would have been cheaper. They make this choice because they do not pay enough attention to the orderbook, or because they have to buy fast from risk management reasons on their clients' orders.
\item High Frequency Participants (HFP) accept a transaction when the imbalance is around one half of the one when Institutional brokers accept a trade. For sure they look more at the orderbook state before taking a decision. Moreover they can probably be more opportunistic: ready to wait the perfect moment instead of being lead by urgency considerations.

If we split HFP between more market making-oriented ones and proprietary trading ones on Figure \ref{fig:imbNASDAQ}.b we see 
\begin{itemize}
\item market makers (probably for inventory reasons: they have to alternate buys and sells), accept to trade when the imbalance is more negative than the average of HFP. They are probably paid back from this adverse selection by bid-ask spread gains (see \cite{citeulike:8423311});
\item proprietary traders are the most opportunistic participants of our panel, leading them to have a less intense imbalance when they trade via limit orders: they seem to be the ones less suffering from adverse selection.
\end{itemize}

\item Global Investment banks are in between. Three reasons may explain such behaviour : first, their activity is a mix of client execution and proprietary trading (hence we perceive the imbalance when they accept a trade as an average of the two categories); second, they have specific strategies to accept transactions via limit orders; third, they invest a little less than HFP in low latency technology, but more than institutional brokers.
\end{itemize}
The \textbf{main effect} to note is each class of agent seems to exploit differently the state of the orderbook before accepting or not a transaction.

\subsubsection{The added value of imbalance for market participants}

Now we know classes of agents take differently into account the state of orderbook imbalance to accept or not a transaction via a limit order, one can ask what could be the value of such a ``\emph{high frequency market timing}''.

We attempt to measure this value with a combination of NASDAQ-OMX labelled transactions and our synchronized market data.
To do this, we compute the midprice move 
 immediately before and after a class of participant $a$ accepts to transact via a limit order:
\begin{equation*}
  \label{eq:midmove}
  \Delta P^{mid}_{\delta t}(\tau,a) = \frac{P^{mid}_{\tau+\delta t} - P^{mid}_\tau}{\bar\psi} \cdot \epsilon_\tau(a),
\end{equation*}
where $\epsilon_\tau(a)$ is the ``sign'' of the transaction (i.e. +1 for a buy and -1 for a sell) and $\bar\psi$ is the average bid-ask spread on the considered stock.

A ``price profile\footnote{Note this ``price profiles'' are now used as a standard way to study the behaviour of high frequency traders in academic papers, see for instance \cite{citeulike:11858957} or \cite{citeulike:13675263}.}'' around a trade is the averaging of this price move as a function of $\delta t$ (between -5 minutes and +5 minutes); it is an estimate of the ``expected price profile'' around a trade:
$$p_a(\delta t) = \frac{1}{\cardT)} \sum_{\tau\in\calT} \Delta P^{mid}_{\delta t}(\tau,a),\; %
\lim_{\cardT)\rightarrow+\infty} p_a(\delta t) = \Esp_\tau \Delta P^{mid}_{\delta t}(\tau,a).$$

\begin{figure}[!ht]
  \centering
  \hfill (a)\hfill \hfill (b)\hfill~\\
  \includegraphics[width=.45\linewidth]{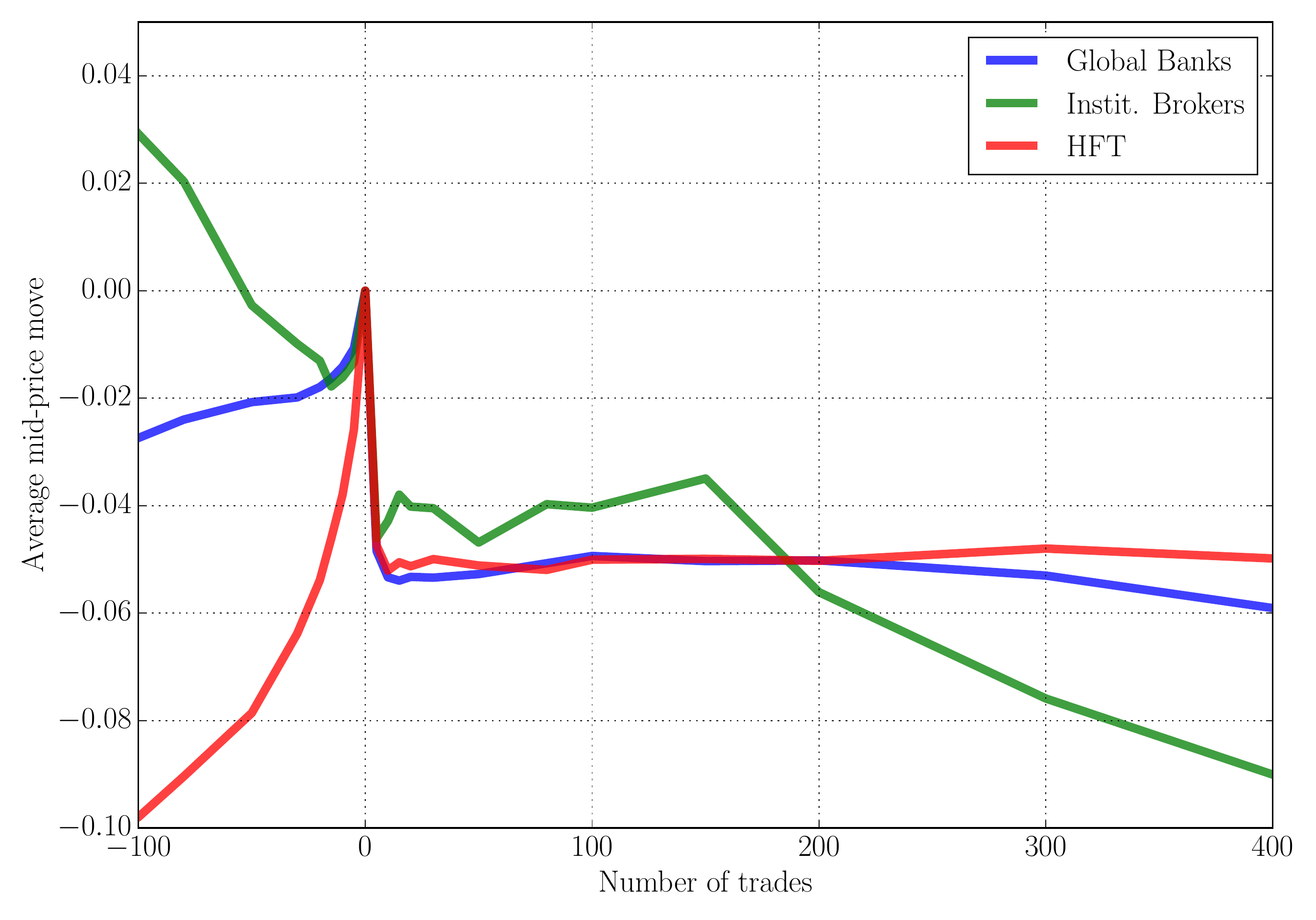}\hfill
  \includegraphics[width=.45\linewidth]{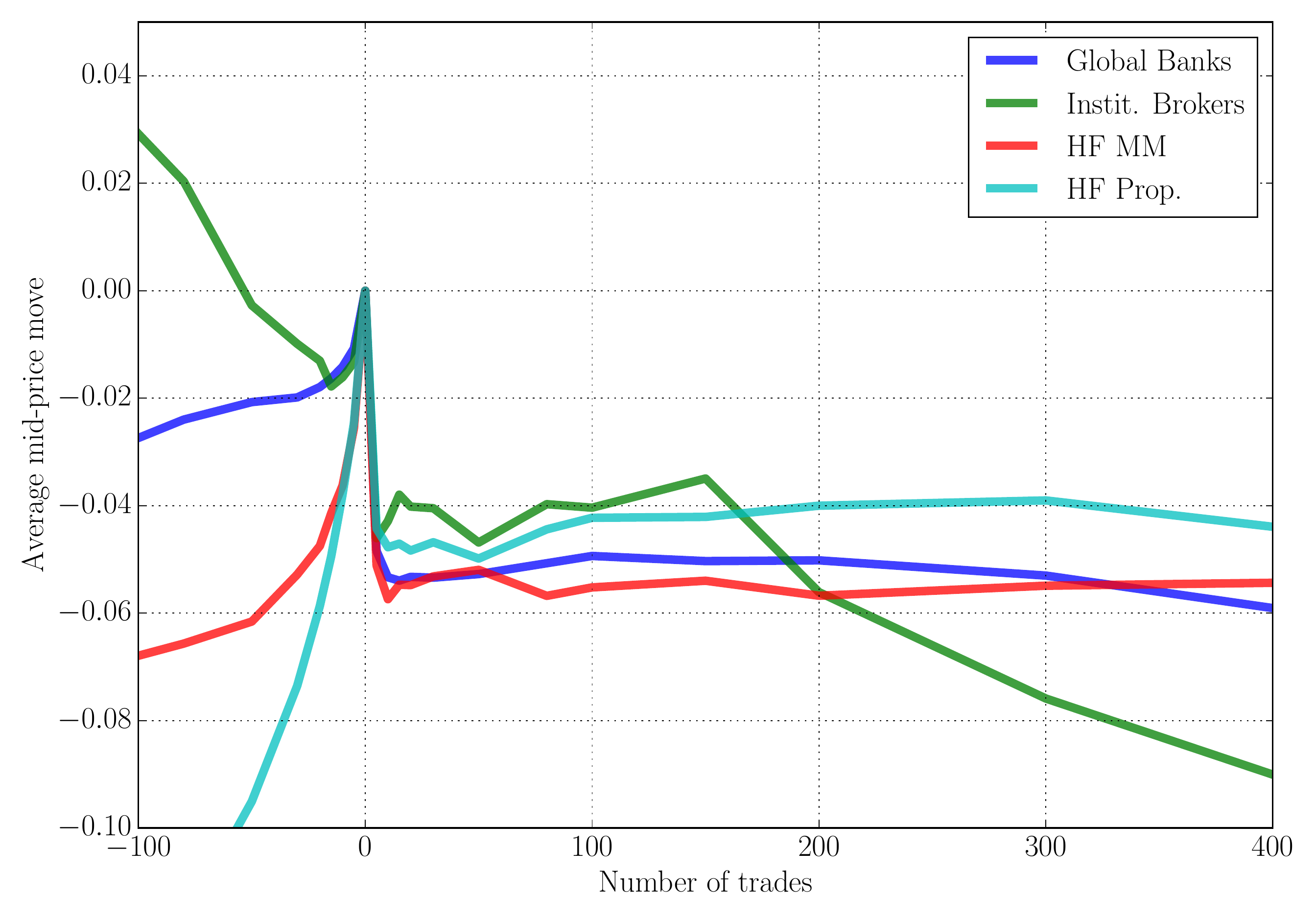}\\
  \caption{Midprice move relative to its position when a limit order is executed for (a) an High Frequency Market Maker, a regional investment bank, and an institutional broker; (b) makes the difference between HF market makers and HF proprietary traders. AstraZeneca (2013-01 to 2013-09).}
  \label{fig:deltaPNASDAQ}
\end{figure}

Figure \ref{fig:deltaPNASDAQ}.a and \ref{fig:deltaPNASDAQ}.b show the price profiles of our three classes of participants, exhibiting real differences beween them. First of all, it confirms the conclusions we draw from Figure \ref{fig:imbNASDAQ}.
Since it is always interesting to have a look at dynamical measures of liquidity (see \cite{citeulike:13616374} for a defense of the use of more dynamical measures of liquidity instead of plain averages):
\begin{itemize}
\item It is clear that Institutional brokers (green line) are buying while the price is going down. Would they have bought later, they would have obtained a cheaper price. As underlined early they probably do it by purpose: they can have urgency reasons or they are using a ``trading benchmark'' that does pay more attention to peg to the executed volume than to the execution price (see \cite[Chapter 3]{citeulike:12047995} for details about brokers' benchmarks).
\item We can see that the difference between High Frequency Participants (HFP) and Global investment banks comes from the price dynamics \emph{before the trade via a limit order}:
for Investment banks the price is more or less stable before the execution and goes down when the limit order is executed. For HFP the price clearly go up \emph{before they bought with a limit order}. It implies that they inserted their limit order shortly before the trade. In our framework we will see how cancelling and reinserting limit orders can be a way to implement an optimal strategy.
\item On Figure \ref{fig:deltaPNASDAQ}.b we see the difference between HF market makers and HF proprietary traders: the latter succeed in inserting buy (resp. sell) limit orders and obtaining transaction while the price is clearly going up (resp. down). After the trade, one can read a difference between them and HF market makers: proprietary traders suffer from less adverse selection (the cyan curve is a little higher than the red one).
\end{itemize}

These charts show that there is a value in taking liquidity imbalance into account.
In Section \ref{sec:understanding}, we show the added of monitoring a limit order by exploiting the orderbook imbalance.


\paragraph{The role of latency.}
Without a fast enough access to the servers of exchanges, a participant could know the best action to perform (insert or cancel a limit order), but not be able to implement it before an unexpected transaction. Since low latency has a cost, some participants may decide to ignore this information and do not access to fast market feeds, orderbooks states, etc.

In the following sections we will not only provide a theoretical framework to ``optimally'' exploit oderbook dynamics for limit order placement, but also study its sensitivity to latency, showing how latency can destroy the added value of understanding orderbook dynamics.

In our theoretical framework, we can explore situations in which the participant knows the best action but cannot implement it on time.

\section{The Dynamic Programming Principle Applied to Limit Order Placement}
\label{sec:dpp}

To setup a discrete time framework for optimal control of one limit order with imbalance, we focus on the simple case (but complex enough in terms of modelling) of one atomic quantity $q_{\epsilon}$ to be executed in $T_f$ units of time (can be orderbook events, trades, or seconds). It will be a buy order, but it is straightforward for the reader to transpose our result for a sell order.

From zero to $T_f$ the trader (or software) in charge of this limit order can : cancel it (i.e. remove it from the orderbook) and insert it at the top of the bid queue (if it is not already in the book), or do nothing. If the trader did not obtain an execution thanks to its optimal posting policy at $T_f$, we force him to cancel his order (if any) and to send a market order to obtain a trade.

For simplicity, we consider a model adapted to large tick stocks (i.e for which the spread equals to one tick). However, our construction can be adapted to a small tick stock by enlarging the control space. For example, we can add the possibility to post a limit order in between the best bid and the best ask. Since we consider a small size order, sending a limit order in between first limits highly increases the adverse selection risk. Consequently, we neglect, as a first approximation, such a control in our model. In a more general framework, other limits should be taken into account and more controls can be considered.
\subsection{Formalisation of the Model}
%

Let $q_{\epsilon}$ be a small limit order inserted at the first bid limit of the orderbook. The orderbook state is modeled by $U^{\mu}_t = \left(Q^{Before,\mu}_t , Q^{After,\mu}_t, Q^{Opp,\mu}_t,P^{\mu}_t \right)$  where $Q^{Before,\mu}$ is the quantity having priority on the order $q_{\epsilon}$, $Q^{After,\mu}$ is the quantity posted after the order $q_{\epsilon}$, $Q^{Opp,\mu}$ is the first opposite limit quantity, $P^{\mu}_t$ is the mid price and $\mu$ is the control of the agent.


For simplicity, we neglect the quantity $q_{\epsilon}$:
\begin{equation*}
 Q^{Same,\mu}= Q^{Before,\mu}+Q^{After,\mu}.
\end{equation*}

\begin{figure}[ht!]
\begin{center}
\begin{tikzpicture}
\draw [double][->](3.5,0) -- ++(0,-0.5);
\draw [double][->](3.5,-5.5) -- ++(0,-0.5);
\draw [double][->](6.5,-1) -- ++(0,-0.5);
\draw [double][->](6.5,-5.5) -- ++(0,-0.5);
\draw (3,-1) rectangle (4,-5);
\draw [black,fill=gray!60] (3,-3) rectangle (4,-3.3);
\draw (6,-2) rectangle (7,-5);
\draw [dotted][->](0,-5) -- ++(10,0);
\draw (5,-5) node[sloped]{$|$};
\draw (3.5,-5) node[below]{$ Same $} ;
\draw (6.5,-5) node[below]{$ Opp $} ;
\draw (3,-4) node[left]{$Q^{Before}$} ;
\draw (3,-3.15) node[left]{$ \text{q} $} ;
\draw (3,-2) node[left]{$Q^{After}$} ;
\draw (7,-4) node[right]{$Q^{Opp}$} ;
\footnotesize
\draw (5,-5) node[below]{$ P(t) $} ;
\normalsize
\draw (10,-5) node[right]{$Price$} ;
\draw (3.5,-0.25) node[right]{$\lambda^{Same,+ }$} ;
\draw (3.5,-5.75) node[right]{$\lambda^{Same,- }$} ;
\draw (6.5,-1.25) node[right]{$\lambda^{Opp,+}$} ;
\draw (6.5,-5.75) node[right]{$\lambda^{Opp,-}$} ;
\end{tikzpicture}
\end{center}  
  \caption{Diagram of flows affecting our orderbook model.}
  \label{fig:lob:flows}
\end{figure}
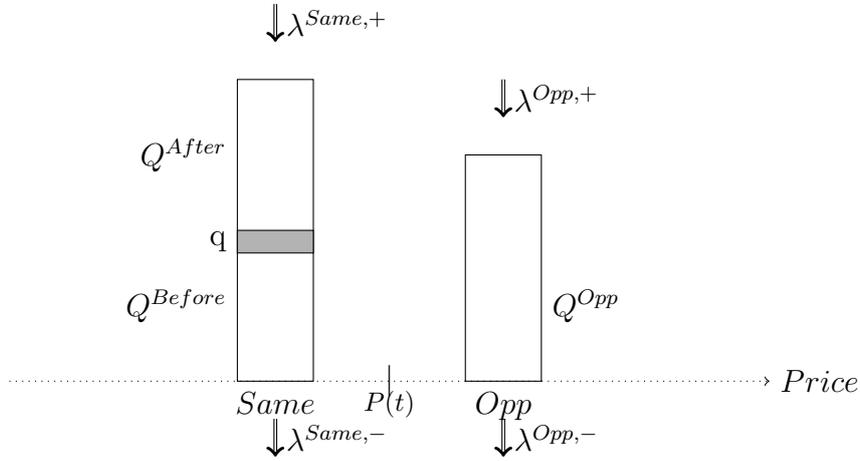

\paragraph{Limit order book dynamics.}
Since we don't differentiate between a cancellation and market orders, the  orderbook dynamics can be modeled by four counting processes (see Figure \ref{fig:lob:flows}):
\begin{itemize}
\item $N_t^{Opp , +}$ (respectively $N_t^{Same , +}$) with an intensity $\lambda^{Opp,+}(Q^{Opp}, Q^{Same})$ (resp. $\lambda^{Same,+}(Q^{Opp}, Q^{Same})$) representing the inserted orders in the opposite limit (resp. same limit).
\item  $N_t^{Opp,-}$ (resp. $N_t^{Same , -}$) with an intensity $\lambda^{Opp,-}(Q^{Opp}, Q^{Same})$ (resp. $\lambda^{Same,-}(Q^{Opp}, Q^{Same})$) representing the canceled orders in the opposite limit (resp. same limit). 
\end{itemize}

In this model, these four counting processes depend only on quantities at first limits. At each event time, an atomic quantity \textbf{q} is added or canceled from the orderbook. Moreover, we assume the bid-ask symmetry relation:
\begin{equation*}
 \left\{
\begin{array}{ccl}
\lambda^{Opp,+}(Q^{Opp}, Q^{Same}) =\lambda^{Same,+}(Q^{Same},Q^{Opp}) \\
\lambda^{Opp,-}(Q^{Opp}, Q^{Same}) =\lambda^{Same,-}(Q^{Same},Q^{Opp}). 
\end{array} \right. 
\end{equation*}

Hence, the size of the first limits can be written as long as none of them turns to be negative :
\begin{equation}
 \left\{
\begin{array}{lcl}
Q^{Opp,\mu}_{t}    & = & Q^{Opp,\mu}_{t^{-}} + q\Delta N_t^{Opp,+} - q\Delta N_t^{Opp,-}  \\
Q^{Before,\mu}_{t} & = & \big(Q^{Before,\mu}_{t^{-}} - q\Delta N_t^{Same,-}\big)\mathrm{1}_{Q^{Before,\mu}_{t^{-}}\geq q}  \\
Q^{After,\mu}_{t}  & = & Q^{After,\mu}_{t^{-}} + q\Delta N_t^{Same,+} - q\mathrm{1}_{0 \leq Q^{Before,\mu}_{t^{-}}< q}\Delta N_t^{Same,-}.
\end{array} \right.
\label{Eq:LOBDyn}
\end{equation}

\subparagraph{What happens when $Q^{After,\mu}$,$Q^{Before,\mu}$ or $Q^{Opp,\mu}$ is totally consumed :} 
First of all, we neglect the probability that at least two of these three events happen simultaneously. When one of the two queues fully depletes, we assume that the price moves in its direction, and we introduce a \emph{discovered quantity} $Q^{Disc}$ to replace the deleted first queue, and an \emph{inserted quantity} $Q^{Ins}$ to be put in front of the opposite one. These quantities are random variables and their law are conditioned by the orderbook state before the depletion. In detail:

\begin{enumerate}
\item {When $ Q^{Opp,\mu}_t = 0 $}. The price increases by one tick (keep in mind for a buy order, the \emph{opposite} is the ask side). 
Then, we discover a new opposite limit and a new bid quantity is inserted into the bid-ask spread (on the bid side) by other market participants (see Figure \ref{fig:lod:down}). It reads
\begin{align*}
 \left\{
\begin{array}{ccl}
    Q^{Opp,\mu}_t    &=& Q^{Disc}(Q^{Opp,\mu}_{t^-},Q^{Same,\mu}_{t^-}) \\
	Q^{Before,\mu}_t &=& Q^{Ins}(Q^{Opp,\mu}_{t^-},Q^{Same,\mu}_{t^-}) \\
	Q^{After,\mu}_t  &=& 0.
\end{array} \right.
\end{align*}
$Q^{Disc}$ is the ``\emph{discovered quantity}'' and $Q^{Ins}$ the ``\emph{inserted quantity}''.

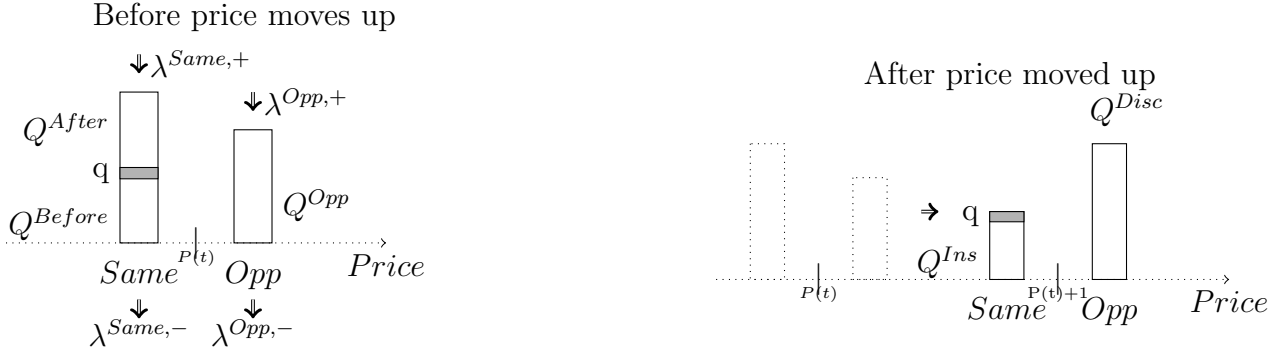
\begin{figure}[ht!]
\begin{minipage}[r]{0.55\textwidth}
\begin{tikzpicture}[scale=0.50]x
\draw [double][->](3.5,0) -- ++(0,-0.5);
\draw [double][->](3.5,-6.5) -- ++(0,-0.5);
\draw [double][->](6.5,-1) -- ++(0,-0.5);
\draw [double][->](6.5,-6.5) -- ++(0,-0.5);
\draw (3,-1) rectangle (4,-5);
\draw [black,fill=gray!60] (3,-3) rectangle (4,-3.3);
\draw (6,-2) rectangle (7,-5);
\draw [dotted][->](0,-5) -- ++(10,0);
\draw (5,-5) node[sloped]{$|$};
\draw (3.5,-5.2) node[below]{$ Same $} ;
\draw (6.5,-5.2) node[below]{$ Opp $} ;
\draw (3,-4.5) node[left]{$Q^{Before}$} ;
\draw (3,-3.15) node[left]{$ \text{q} $} ;
\draw (3,-2) node[left]{$Q^{After}$} ;
\draw (7,-4) node[right]{$Q^{Opp}$} ;
\tiny
\draw (5,-5) node[below]{$ P(t) $} ;
\normalsize
\draw (10,-5) node[below]{$Price$} ;
\draw (3.5,-0.25) node[right]{$\lambda^{Same,+}$} ;
\draw (3.5,-6.75) node[below]{$\lambda^{Same,-}$} ;
\draw (6.5,-1.25) node[right]{$\lambda^{Opp,+}$} ;
\draw (6.5,-6.75) node[below]{$\lambda^{Opp,-}$} ;
\draw (2,1) node[right]{Before price moves up} ;
\end{tikzpicture}\\
\end{minipage} \hfill
\begin{minipage}[l]{0.55\textwidth}
\begin{tikzpicture}[scale=0.45]
\draw [double][->](6,-3) -- ++(0.5,0);
\draw [dotted](1,-1) rectangle (2,-5);
\draw (8,-3) rectangle (9,-5);
\draw [black,fill=gray!60] (8,-3) rectangle (9,-3.3);
\draw [dotted](4,-2) rectangle (5,-5);
\draw (11,-1) rectangle (12,-5);
\draw [dotted][->](0,-5) -- ++(15,0);
\draw (3,-5) node[sloped]{$|$};
\draw (10,-5) node[sloped]{$|$};
\draw (8.5,-5.2) node[below]{$ Same $} ;
\draw (11.5,-5.2) node[below]{$ Opp $} ;
\draw (8,-4.5) node[left]{$Q^{Ins}$} ;
\draw (8,-3.15) node[left]{$ \text{q} $} ;
\draw (12,0) node{$Q^{Disc}$} ;
\tiny
\draw (3,-5) node[below]{$ P(t) $} ;
\draw (10,-5) node[below]{P(t)+1} ;
\normalsize
\draw (15,-5) node[below]{$Price$} ;
\draw (4,1) node[right]{After price moved up} ;
\end{tikzpicture}
\end{minipage}  
  \caption{Diagram of a upward price change for our model.}
  \label{fig:lod:down}
\end{figure}

\item {When $ Q^{Before,\mu}_t = 0 $}. The limit order is executed. This case is considered in equation  \ref{Eq:LOBDyn}.

\item {When moreover $ Q^{After,\mu}_t = 0 $}. The price decreases by one tick. Then, we discover a new quantity on the bid side and market makers insert a new quantity on the opposite side :
\begin{align*}
 \left\{
\begin{array}{ccl}
Q^{Opp,\mu}_t    &=& Q^{Ins}(Q^{Opp,\mu}_{t^-},Q^{Same,\mu}_{t^-}) \\
Q^{Before,\mu}_t &=& Q^{Disc}(Q^{Opp,\mu}_{t^-},Q^{Same,\mu}_{t^-}) \\
Q^{After,\mu}_t  &=& 0
\end{array} \right.
\end{align*}
If the limit order was in the orderbook: it has been executed. Otherwise the price moves down and the trader has the opportunity to reinsert a limit order on the top of $Q^{Disc}$ (see Figure \ref{fig:lob:model:upward} for a diagram).

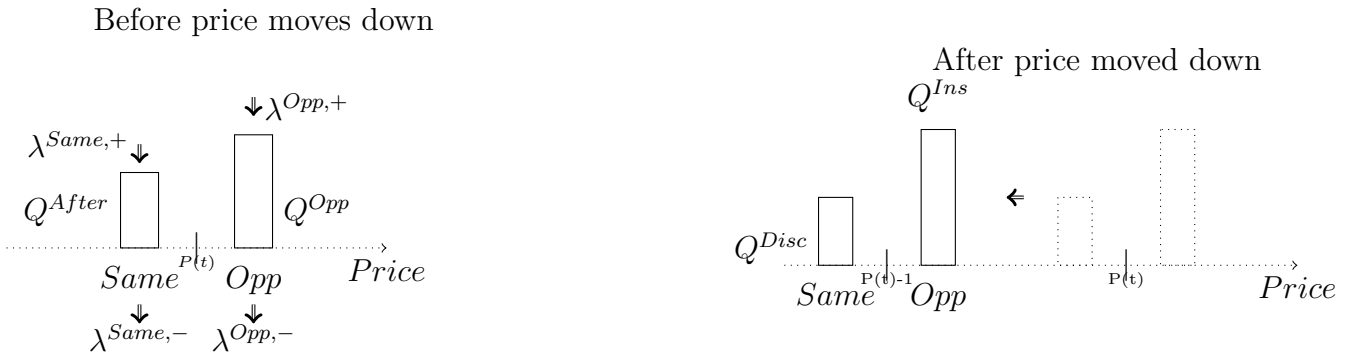
\begin{figure}[ht!]
  \begin{minipage}[c]{0.55\linewidth}
\begin{tikzpicture}[scale=0.50]
\draw [double][->](3.5,-2.25) -- ++(0,-0.5);
\draw [double][->](3.5,-6.5) -- ++(0,-0.5);
\draw [double][->](6.5,-1) -- ++(0,-0.5);
\draw [double][->](6.5,-6.5) -- ++(0,-0.5);
\draw (3,-3) rectangle (4,-5);
\draw (6,-2) rectangle (7,-5);
\draw [dotted][->](0,-5) -- ++(10,0);
\draw (5,-5) node[sloped]{$|$};
\draw (3.5,-5.2) node[below]{$ Same $} ;
\draw (6.5,-5.2) node[below]{$ Opp $} ;
\draw (3,-4) node[left]{$Q^{After}$} ;
\draw (7,-4) node[right]{$Q^{Opp}$} ;
\tiny
\draw (5,-5) node[below]{$ P(t) $} ;
\normalsize
\draw (10,-5) node[below]{$Price$} ;
\draw (3.5,-2.25) node[left]{$\lambda^{Same,+}$} ;
\draw (3.5,-6.75) node[below]{$\lambda^{Same,-}$} ;
\draw (6.5,-1.25) node[right]{$\lambda^{Opp,+}$} ;
\draw (6.5,-6.75) node[below]{$\lambda^{Opp,-}$} ;
\draw (2,1) node[right]{Before price moves down} ;
\end{tikzpicture}\\

\end{minipage} \hfill
\begin{minipage}[c]{1\linewidth}
\begin{tikzpicture}[scale=0.45]
\draw [double][->](7,-3) -- ++(-0.5,0);
\draw [dotted](8,-3) rectangle (9,-5);
\draw (1,-3) rectangle (2,-5);
\draw [dotted](11,-1) rectangle (12,-5);
\draw (4,-1) rectangle (5,-5);
\draw [dotted][->](0,-5) -- ++(15,0);
\draw (3,-5) node[sloped]{$|$};
\draw (10,-5) node[sloped]{$|$};
\draw (1.5,-5.2) node[below]{$ Same $} ;
\draw (4.5,-5.2) node[below]{$ Opp $} ;
\draw (1,-4.5) node[left]{$Q^{Disc}$} ;
\draw (4.5,0) node{$Q^{Ins}$} ;
\tiny
\draw (3,-5) node[below]{$ \text{P(t)-1}$} ;
\draw (10,-5) node[below]{P(t)} ;
\normalsize
\draw (15,-5) node[below]{$Price$} ;
\draw (4,1) node[right]{After price moved down} ;
\end{tikzpicture}\\
\end{minipage}
  \caption{Diagram of an downward price move in our model.}
  \label{fig:lob:model:upward}
\end{figure}
\end{enumerate}

\subparagraph{The control.} 
We consider two types of control $ C = \left\{s ,c \right\} $:
\begin{itemize}
\item $c$ (like \emph{continue}): stay in the orderbook.
\item $s$ (like \emph{stop}): cancel the order  and wait for a better orderbook state to reinsert it at the top of $Q^{same}$ ($Q^{bid}$ for our buy order). This control is essentially used to avoid adverse selection, i.e. avoid to buy just before a price decrease.
\end{itemize}

\paragraph{Optimal control problem.} We fix a finite time horizon $T_f<\infty$ and we want to compute : 
\begin{equation*}
V_{T}(0,U) = \underset{\mu}{\sup}\, \Esp \big[  \Delta P^{\mu}_{\infty} \big].
\end{equation*}
Where : 
\begin{itemize}
\item $U = (q^{before},q^{after},q^{opp},p)$ is the initial state of the orderbook. 
\item $T^{\mu}_{Exec} = \inf\big\{t \geq 0, \, s.t \, Q^{Before,\mu}_t < q, \, \mu_t = c\big\}\wedge T_f$ represents the first execution time. Once the order executed, the orderbook is frozen.
\item $\Delta P^{\mu}_{\infty} = \underset{t \rightarrow \infty}{\lim} \big(P^{\mu}_{t} - P^{Exec,\mu}_{T^{\mu}_{Exec}} \big) $ represents the gain of the trader, where the execution price $P^{Exec,\mu}_{t}$ satisfy $P^{Exec,\mu}_{t} = P^{\mu}_{t} - \frac{1}{2}$ when the limit order is executed before $T_f$ and $P^{Exec,\mu}_{t} = P^{\mu}_{t} + \frac{1}{2}$ otherwise. Indeed, if at $T_f$ the order hasn't been executed, we cross the spread to guarantee execution.   
\end{itemize}

\paragraph{Choice of a benchmark.}
We compare the value of the obtained shares at $t$ to \emph{its expected value at infinity}, i.e. $\Esp (P_{+\infty}(t) | \Imb_t) $ since it is not attractive to buy at the best bid if we expect the price to continue to go down. Indeed, it is possible to expect a better future price thanks to the observed imbalance. It thus induces an \emph{adverse selection cost} in the framework.

This is not a detail since the \emph{trader will have no insentive to put a limit order at the top of a very small queue if the opposite side of the book is large}. We will see in Section \ref{sec:understanding} using empirical evidences that this is a realistic behaviour. Such a behaviour cannot be captured by other linear frameworks like \cite{moa14hjb}.

To solve the control problem, we introduce in the next section a discrete time version of the initial problem whose value function can be computed numerically. In this discrete framework, $\underset{t \rightarrow \infty}{\lim} \big(P^{\mu}_{t} - P^{Exec,\mu}_{T^{\mu}_{Exec}} \big) $ is computed using the imbalance.

\subsection{Discrete time framework}


We set the time step to $\Delta_t$ and the final time to $T_{f}$. Let $t_0=0 < t_1 \cdots t_{f-1} < t_{f}= T_f$ different instants at which the orderbook is observed, such that $ t_n = n\Delta_t $ for all $ n \in \left\{0 ,1, \cdots ,f \right \} $.

Under the assumption that between two consecutive instants $t_n$ and $t_{n+1}$ for all $ n \in \left\{0 ,1, \cdots ,f-1 \right \} $, only five cases can occur:
\begin{itemize}
\item 1 unit quantity is added at the bid side;
\item 1 unit quantity is consumed at the bid side;
\item 1 unit quantity is added at the opposite side;
\item 1 unit quantity is consumed at the opposite side;
\item nothing happens.
\end{itemize}

We neglect the situation where at least two cases occur during the same time interval (the probability of such conjunctions are of the orders of $\lambda^2$, hence our approximation remains valid as far as $\lambda^2 dt$ is small compared to $\lambda dt$).

\begin{framework}[Our setup in few words.]
  In short, our main assumptions are:
  \begin{itemize}
  \item only one limit order of small quantity $\textbf{q}_{\epsilon}$ is controlled, it is small enough to have no influence on orderbook imbalance;
  \item decrease of queue sizes at first limits is caused by transactions only (i.e. no difference between cancellation and trades);
  \item queues decrease or increase by one quantity only;
  \item the intensities of point processes (including the ones driving quantities inserted into the bid-ask spread, and driving the quantity discovered when a second limit becomes a first limit) are functions of the quantities at best limits only;
  \item no notable conjunction of multiple events.
  \end{itemize}
\end{framework}

We introduce the following Markov chain $ U^{\mu}_n = \left( Q^{Before,\mu}_n, Q^{After,\mu}_n, Q^{ Opp,\mu}_n, P^{\mu}_n , \text{Exec}_n \right)$ where: 
\begin{itemize}
\item $Q^{Before,\mu}_n$ is the  $Q^{Before,\mu}$  size at time $t_n$ that takes value in $\mathbb{N}$.
\item $Q^{After,\mu}_n$ is the $Q^{After,\mu}$ size at time $t_n$ that takes value in $\mathbb{N} $.
\item $Q^{Opp,\mu}_n$ is the $Q^{Opp,\mu}$ size at time $t_n$ that takes value in $\mathbb{N}$.
\item $P^{\mu}_n$ is the mid price at time $t_n$. 
\item $\text{Exec}_n$ is an additional variable taking values in $ \{-1,0,1 \} $. $\text{Exec}_n$ equals to 1 when the order is executed at time $t_n$, 0 when the order is not executed at time $t_n$ and -1 (a ``\emph{cemetery state}'') when the order has been already executed before $t_n$.
We set $ \text{Exec}_0=0 $.
\end{itemize}
In the same way, we define $N^{Same,+}_{n}$, $N^{Same,-}_{n} $, $N^{Opp,+}_{n} $  and $N^{Opp,-}_{n}$ as the values of the counting processes $N^{Same,+}_{t}$, $N^{Same,-}_{t} $, $N^{Opp,+}_{t} $  et $N^{Opp,-}_{t}$ at time $t_n$. The transition probabilities of the markov chain $U_n$ are detailed in Appendix \ref{sec:app:transition}.

\paragraph{The terminal constraint.}
The microprice $P_{\infty,k} \approx \Esp (P_{\infty}|\mathcal{F}_k) $ is defined such as:
\begin{equation*}
P_{\infty,k} =F(Q^{Opp}_{k},Q^{Same}_{k},P_{k}) = P_{k} + \frac{\alpha}{2} \cdot \frac{Q^{Same}_{k} - Q^{Opp}_{k} }{ Q^{Opp}_{k}+Q^{Same}_{k} } \quad  \forall k \in \{0,1, \cdots , f \}.
\label{Eq:TermConstr1}
\end{equation*}
Where $\mathcal{F}_k$ is the filtration associated to $U_k$ such that $\mathcal{F}_k = \sigma \left(U_n , n\leq k \right)$ and $\alpha$ is a parameter that represents the sensitivity of futur prices to the imbalance.\\ 
The execution price $P_{\text{Exec},k} $ is defined $\forall k \in \{0,1, \cdots , f \} $ such that :
\begin{equation*}
P_{\text{Exec},k} =  \left\{
\begin{array}{l}
    P_{k} + \frac{1}{2} \qquad \text{when} \; \text{Exec}_k = 0\\
    P_{k} + \frac{1}{2} \qquad \text{when}\; \text{Exec}_k \in \left\{-1,1\right\}\, \text{and}\, P_{k+1} - P_k \ne 0 \\
	P_{k} - \frac{1}{2} \qquad \text{when}\; \text{Exec}_k \in \left\{-1,1\right\}\, \text{and}\, P_{k+1} - P_k = 0.
\end{array} \right.
\end{equation*}

Let $k_0$ be the execution time: $k_0 = \inf \left(k \geq 0 , \text{Exec}_k = 1 \right) \wedge f$. Then, the terminal valuation can be written: 
\begin{equation}
Z_{k_0} = P_{\infty,k_0} - P_{\text{Exec},k_0} 
\label{Eq:FinalConst02}
\end{equation}

Let $\mathcal{U}$ the set of all progressively measurable processes $ \mu := \left\{\mu_k , k < f \right\}$ valued in $ \left\{s,c\right\} $
This problem can be written as a stochastic control problem : 
\begin{equation*}
V_{U_0,f} = \underset{\mu \in \mathcal{U}}{\sup} \underset{U_0,\mu}{\Esp} \left( Z_{k_0}\right)  = \underset{\mu \in \mathcal{U}}{\sup} \underset{U_0,\mu}{\Esp} \left( \sum_{i=1}^{f-1} g_i(U_i,\mu_i) + g_f(U_f)\right).
\end{equation*}
Where $ g_i(U_i,\mu_i) = Z_{i}$ when $\text{Exec}_i =1 $ and $\mu_i = c $ and 0 otherwise for all $i \in \left\{1,\cdots,f-1 \right\}$, and $g_f(U_f) = Z_f$ when $\text{Exec}_f \in  \left\{ 0,1\right\} $ and 0 otherwise.

We want to compute $V_{U_0,f}= \underset{\mu \in \mathcal{U}}{\sup}  \, \underset{U_0,\mu}{\Esp}(Z_{k_0})$ using dynamic programming algorithm:  
\begin{equation}
\left\{ \begin{array}{l}
G_f = Z_f \\
G_n = \max \left( P^{c}_{n}G_{n+1} , P^{s}_{n}G_{n+1} \right) \quad  \forall \, n \in \{0,1, \cdots , f-1 \}.
\end{array} \right.
\label{eq:loigenerale}
\end{equation}
Where $P_n$ represents the transition matrix of the markov chain $U_n$.


%
\section{A Qualitative Understanding}
\label{sec:understanding}

Equation (\ref{eq:loigenerale}) provides an explicit forward-backward algorithm that can be solved numerically:
\begin{itemize}
\item \emph{Step 1 Forward simulation :} Starting from an initial state $u$, we simulate all the reachable states during $f$ periods.
\item \emph{Step 2 Backward simulation:} At the final period $f$, we can compute $G_f$ for each reachable state. Then, using the backward equation (\ref{eq:loigenerale}), we can compute, recursively, $G_i$ knowing $G_{i+1}$ to get $G_0$.
\end{itemize}
 In this section, we present and comment the simulation results. For more details about the forward-backward algorithm see Appendix \ref{sec:app:transition}. 

We are going to compare two situations:
\begin{itemize}
\item The first one called   \textbf{(NC)} \manuallabel{txt:NC}{NC}  corresponds  to the case  when 
  no control is adopted (i.e we always stay in the orderbook and ``join the best bid'' each time it changes). 
\item  The second one called \textbf{(OC)} \manuallabel{txt:OC}{OC} corresponds  to the optimal control case: controls "c" and "s" are considered. 
\end{itemize}
Moreover, our simulation results are given for two different cases :
\begin{itemize}
\item \textbf{Framework (CONST):} \manuallabel{txt:CONST}{CONST} intensities of insertion and cancellation are constant: $\lambda^{Same,+}_k =  \lambda^{Opp,+}_k = 0.06 $ and $\lambda^{Same,-}_k =  \lambda^{Opp,-}_k = 0.5 \,$  $\forall k \in \left\{0,1,\cdots,f\right\}$. Under (\ref{txt:CONST}), the inserted quantities $Q^{Ins}$ and discovered quantities $Q^{Disc}$ are constant too.
\item \textbf{Framework (IMB):} \manuallabel{txt:IMB}{IMB} intensities of cancellation and insertion are functions of the imbalance such as $\forall k \in \left\{0,1,\cdots,f\right\}$ : 

\begin{equation*}
\begin{array}{l}
\lambda^{Opp,+}_k\left(Q^{Opp}_k,Q^{Same}_k\right)= \lambda^{Same,+}_k\left(Q^{Same}_k,Q^{Opp}_k\right) = \lambda^{+}_0 + \beta^{+}\frac{Q^{Opp}_k}{(Q^{Opp}_k+Q^{Same}_k)} \\
\lambda^{Opp,-}_k\left(Q^{Opp}_k,Q^{Same}_k\right)= \lambda^{Same,-}_k\left(Q^{Same}_k,Q^{Opp}_k\right) = \lambda^{-}_0 + \beta^{-}\frac{Q^{Same}_k}{ (Q^{Opp}_k+Q^{Same}_k)} \\
\end{array}
\end{equation*}
Where $\lambda^{\pm}_0$ are basic insertion and cancellation intensties and $\beta^{\pm}$ are predictability parameters representing the sensitivity of order flows to the imbalance.

Moreover, under (\ref{txt:IMB}), inserted and discovered quantities are computed in the following way:

\begin{itemize}
\item \textbf{When $Q^{Opp}_k$ is totally consumed}, we set $Q^{Disc}_k = \lceil q^{disc}_0 + \theta_{disc} \cdot Q^{Same}_k \rceil $ and $Q^{Ins} = \lceil q^{ins}_0 + \theta_{ins} \cdot Q^{Same}_k \rceil$. Where $\theta_{disc}$ and $\theta_{ins}$ are coefficients associated to liquidity 
and $\lceil . \rceil $ is the upper rounding. $q^{disc}_0$ and $q^{ins}_0$ are the basic discovered and inserted quantities. 
\item Similarly \textbf{when $Q^{Same}_k$ is totally consumed}, we set $Q^{Disc} = \lceil q^{disc}_0 + \theta_{disc} \cdot Q^{Opp}_k \rceil $ and $Q^{Ins} = \lceil q^{ins}_0 + \theta_{ins} \cdot Q^{Opp}_k \rceil $.
\end{itemize}
\end{itemize}
This kind of relations is compatible with empirical findings of \cite{citeulike:12810809} and different from \cite{citeulike:8531765} in which $Q^{Disc}= Q^{Ins}$ is independant of liquidity imbalance.

\subsection{Numerically Solving the Control Problem}

\subsubsection{Anticipation of Adverse Selection}

The cancellation is used by the optimal strategy to
avoid adverse selection. For instance, when the quantity on the same side is extremely lower than the one on the opposite side, it is expected to cancel the order to wait for a better future opportunity. The optimal control takes in consideration this effect and cancels the order when such a high adverse selection effect is present. 

We keep notations of Section \ref{sec:dpp}. Let $\mu := \left\{ \mu_k, k<f \right\}$ a control, we define $ \Esp_{U_0,\mu} \left( \Delta \text{P} | \text{Exec} \right) = \Esp_{U_0,\mu} \left( Z_{k_0} \right) $. $ \Esp_{U_0,\mu} \left( \Delta \text{P} | \text{Exec}\right)$ depends on the control $\mu$, the initial state of the orderbook $U_0$ and the terminal period $f$. The quantity $\Esp_{U_0,\mu} \left( \Delta \text{P} | \text{Exec} \right)$ can be directly computed by a forward algorithm that visits all the possible states of the markov chain $U^{\mu}_n$. For more details about the transition probabilities of the Markov chain $U^{\mu}_n$ see Appendix \ref{sec:app:transition}.

The quantity $\Esp_{U_0,\mu} \left( \Delta \text{P} | \text{Exec} \right) $ is interesting since it corresponds to the quantity to maximize in our optimal control problem and it represents as well the profitability/trade of an agent. 

\begin{figure}[!ht]
	\centering
    \hfill (a)\hfill (b) \hfill~\\
	\includegraphics[width=.45\linewidth]{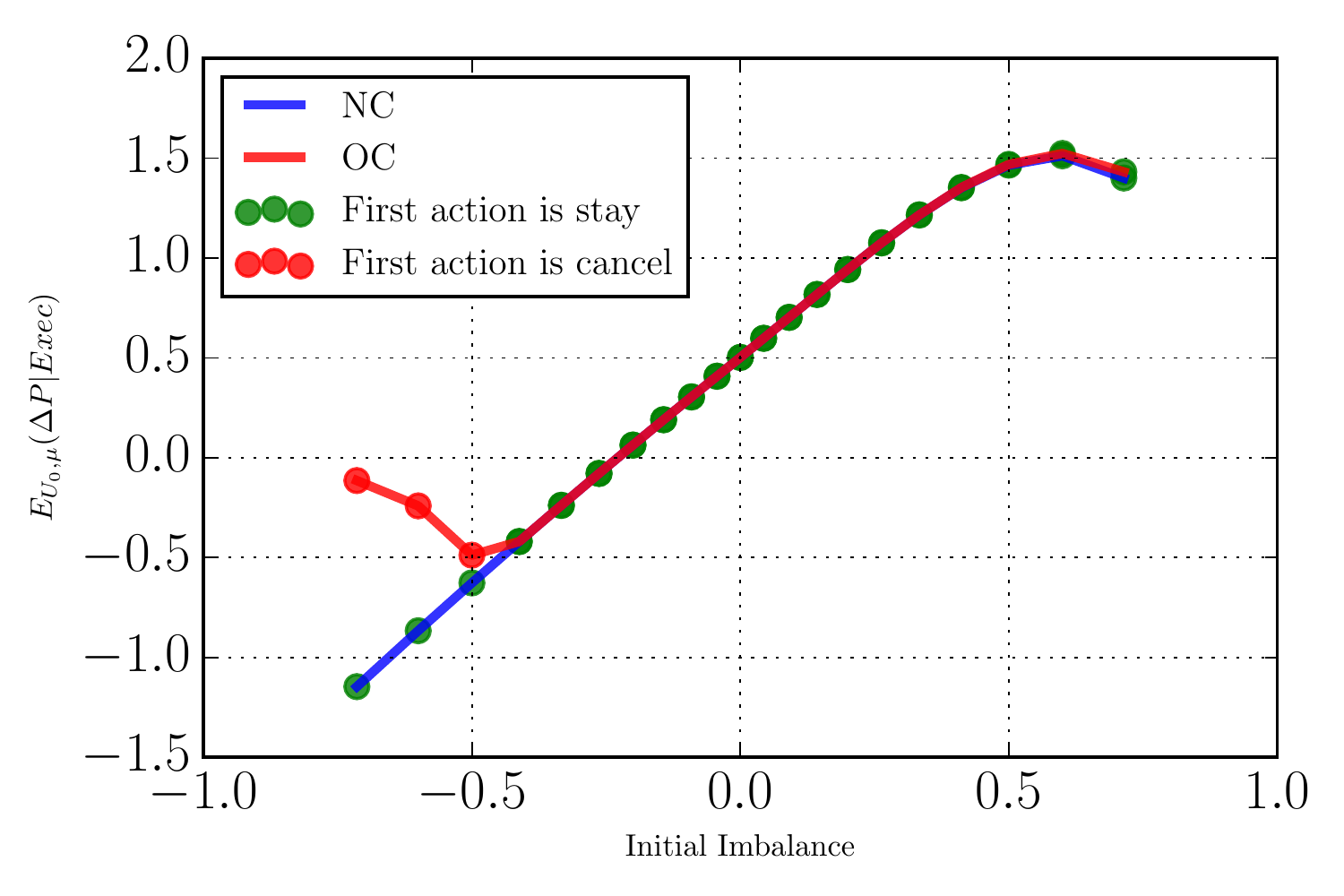}
	\includegraphics[width=.45\linewidth]{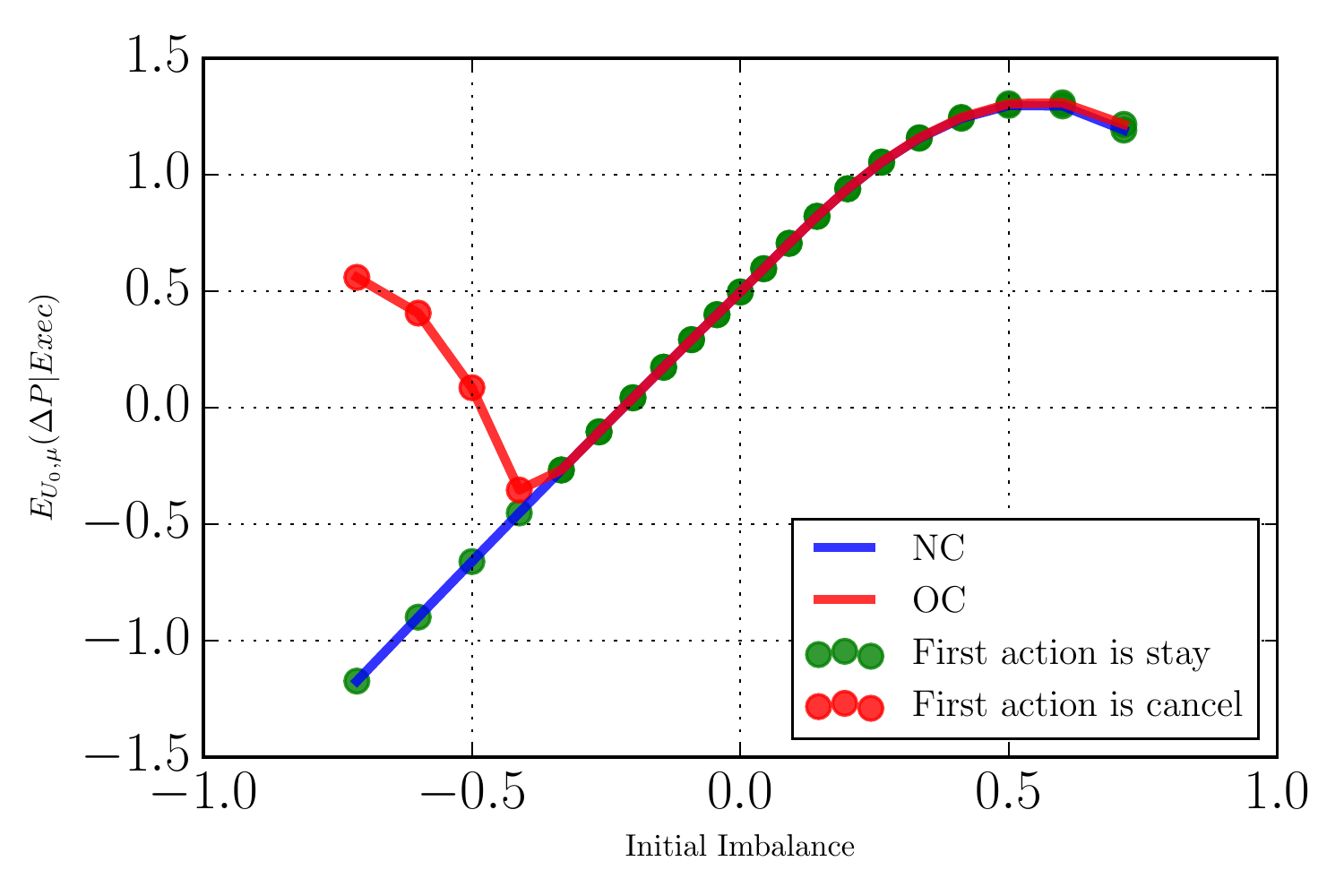} 
  \caption{(a) (resp. (b)) $\Esp_{U_0,\mu^*} \left( \Delta \text{P} | \text{Exec} \right)$  and $\Esp_{U_0,\mu^c} \left( \Delta \text{P} | \text{Exec} \right)$ when intensities are constant (\ref{txt:CONST}) (resp . (\ref{txt:IMB})).}
  \label{fig:deltaPImbalanceNC_Vs_OC}
\end{figure}

Let $\mu^c$ the control where the agent always stays in the orderbook (i.e \ref{txt:NC}) and $\mu^*$ the optimal control (i.e \ref{txt:OC}). The Figure \ref{fig:deltaPImbalanceNC_Vs_OC}.a represents the variation of $\Esp_{U_0,\mu^*} \left( \Delta \text{P} | \text{Exec} \right)$ and $\Esp_{U_0,\mu^c} \left( \Delta \text{P} | \text{Exec} \right)$ when the initial imbalance of the orderbook moves under (\ref{txt:CONST}). In Figure \ref{fig:deltaPImbalanceNC_Vs_OC}.a, blue points are initial states where it is optimal to stay in the orderbook since the beginning (i.e $t=0$) while red points are initial states where it is optimal to cancel the order at $t=0$. Initial parameters are fixed such that $\lambda^{Same,+} = \lambda^{Opp,+} = 0.06 $, $\lambda^{Same,-} =  \lambda^{Opp,-} = 0.5$, $\alpha =4$, $Q^{Disc} =6$, $Q^{Ins} = 4$, $f=20$, $q=1$ and $P_0 = 10$. Moreover, initial imbalance values are obtained by varying $Q^{Opp}_0$ from 2 to 12 and $Q^{After}_0$ from 1 to 11 while $Q^{Before}_0$ is kept constant equal to 1.

Figure \ref{fig:deltaPImbalanceNC_Vs_OC}.b is the analogous of Figure \ref{fig:deltaPImbalanceNC_Vs_OC}.a but under the framework (\ref{txt:IMB}). In Figure \ref{fig:deltaPImbalanceNC_Vs_OC}.b, initial parameters are fixed such that $\lambda^{+}_0 = 0.06$, $\lambda^{-}_0 = 0.5$, $\beta^{+} = 0.075$, $\beta^{-} = 0.25$, $q^{disc}_0 = 6$, $q^{ins}_0 = 2$, $\theta_{disc} = 3$, $\theta_{disc} = 0.5$, $\alpha =4$, $f=20$ and $P_0 = 10$. Similarly, initial imbalance values are obtained by varying $Q^{Opp}_0$ from 2 to 12 and $Q^{After}_0$ from 1 to 11 while $Q^{Before}_0$ is kept constant equal to 1.
 
The \textbf{main effect} to note on these curves is the way the optimal control anticipates adverse selection. When imbalance is highly negative, we cancel first the order (red points) to take advantage from a better futur opportunity. We notice that under the framework (\ref{txt:IMB}), the agent cancels earlier than under the framework (\ref{txt:CONST}) since more weights are given to cancellation events. This point is detailed in \ref{IMBIntensInf}.

Appendix \ref{sec:extremb} explains the downward slopes at the left of Figures \ref{fig:deltaPImbalanceNC_Vs_OC}.a and \ref{fig:deltaPImbalanceNC_Vs_OC}.b.

\subsubsection{Price Improvement comes from avoiding adverse selection}

As expected, results obtained in the optimal control (\ref{txt:OC}) case are better than the ones in the non-controlled (\ref{txt:NC}) case : by cancelling and taking into account liquidity imbalance, one can be more efficient than just staying in the orderbook.

\begin{figure}[!ht]
  \centering
  \hfill (a)\hfill (b) \hfill~\\
  \includegraphics[width=.45\linewidth]{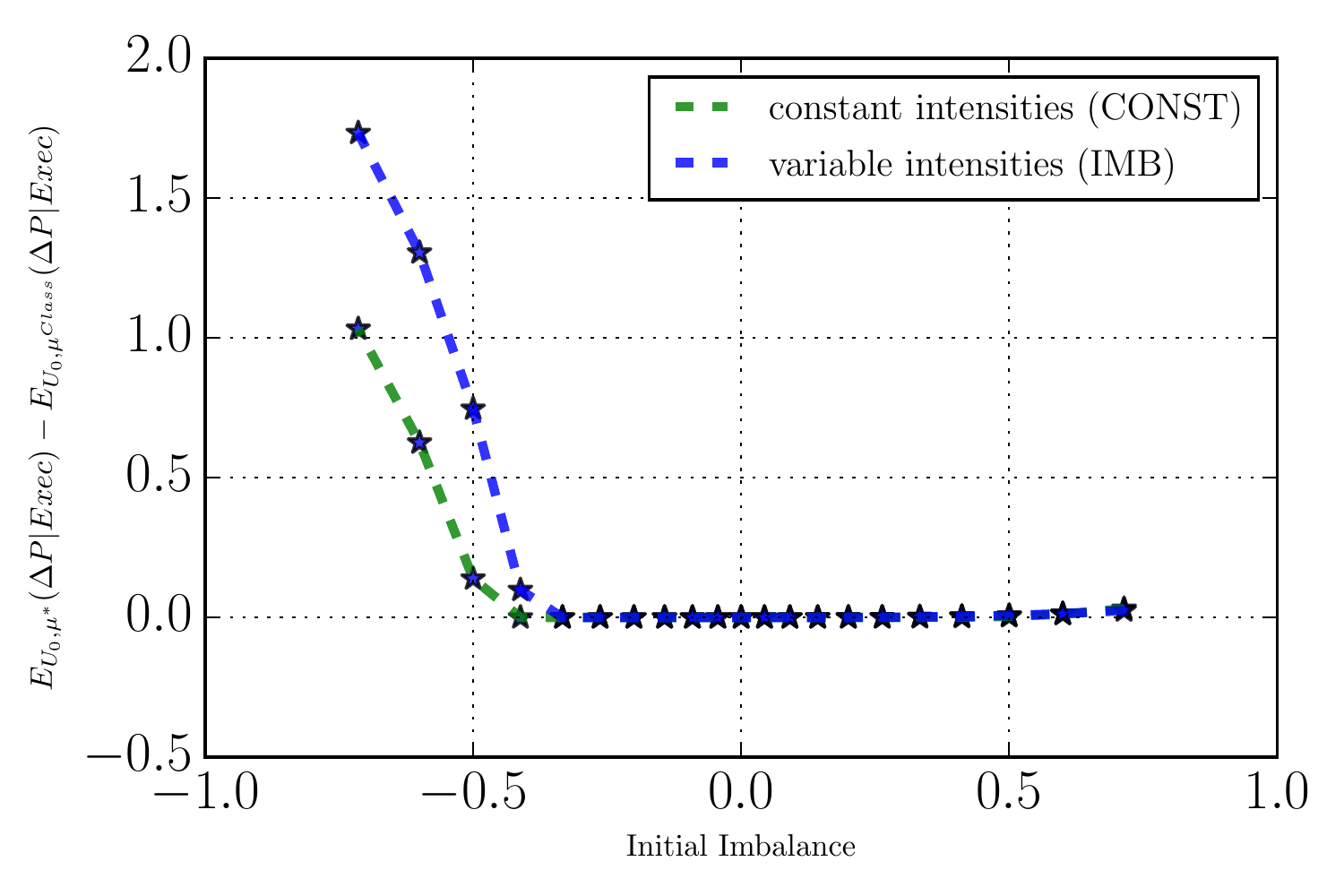}
  \includegraphics[width=.45\linewidth]{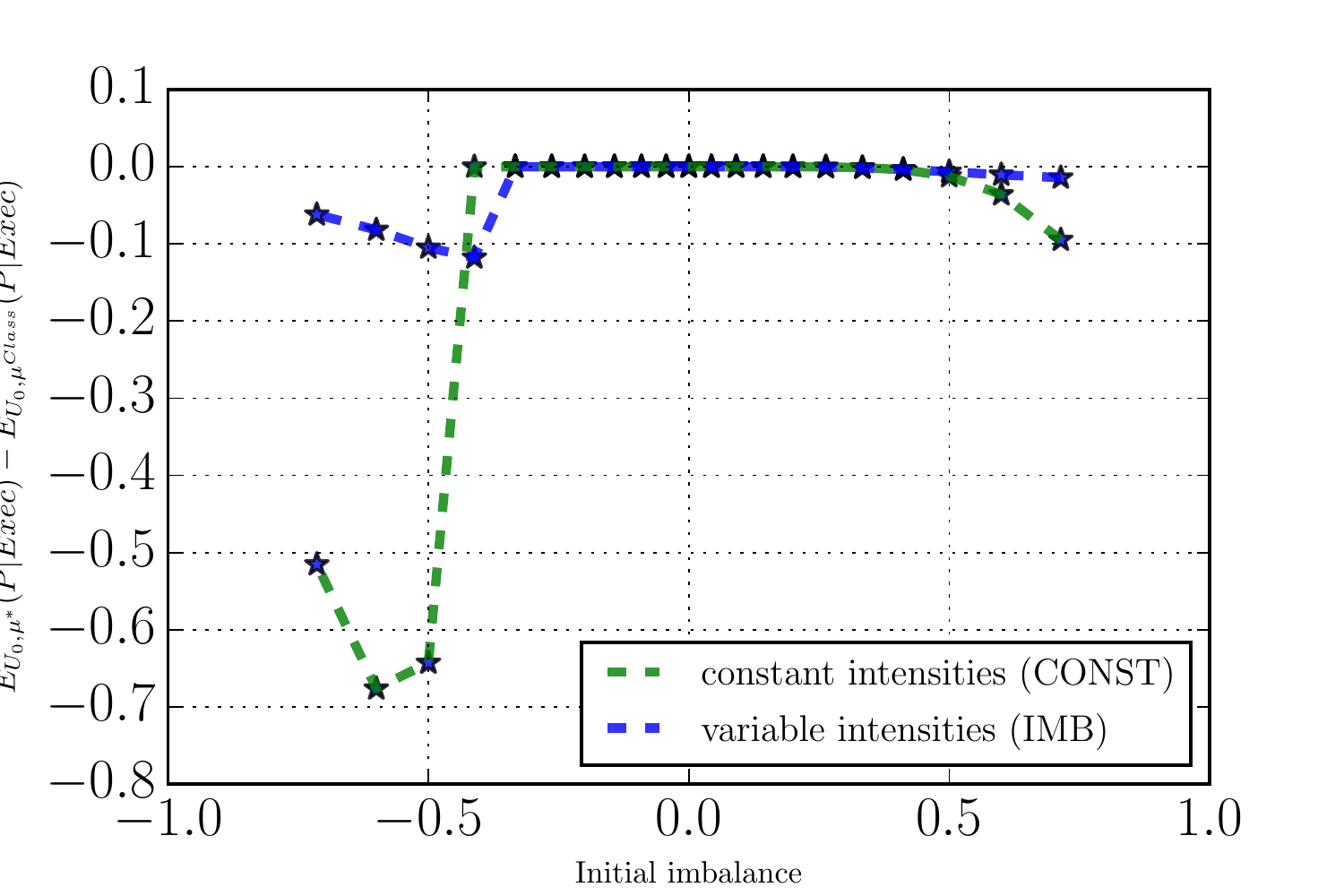}
  \caption{ (a) $\Esp_{U_0,\mu^*} \left( \Delta \text{P} | \text{Exec} \right) - \Esp_{U_0,\mu^c} \left( \Delta \text{P} | \text{Exec} \right)$ move relative to initial imbalance under (\ref{txt:CONST}) and (\ref{txt:IMB}). (b)  $\Esp_{U_0,\mu^*} \left(P_{k_0} | \text{Exec} \right) - \Esp_{U_0,\mu^c} \left( P_{k_0} | \text{Exec} \right)$ move relative to initial imbalance move relative to initial imbalance under (\ref{txt:CONST}) and (\ref{txt:IMB}). }
  \label{fig:deltaPErrorImbalance}
\end{figure}

Figure \ref{fig:deltaPErrorImbalance}.a shows the variation of the price improvement (resp. $\Esp_{U_0,\mu^*} \left( \Delta \text{P} | \text{Exec} \right) - \Esp_{U_0,\mu^c} \left( \Delta \text{P} | \text{Exec} \right)$) when the initial imbalance moves, under both frameworks (\ref{txt:CONST}) and (\ref{txt:IMB}). We kept the same initial parameters of Figures \ref{fig:deltaPImbalanceNC_Vs_OC}.a and \ref{fig:deltaPImbalanceNC_Vs_OC}.b.

Similarly, Figure \ref{fig:deltaPErrorImbalance}.b represents the variation $ \Esp_{U_0,\mu^*} \left( P_{k_0} | \text{Exec} \right) - \Esp_{U_0,\mu^c} \left(P_{k_0} | \text{Exec} \right)$  when the initial imbalance moves, under both frameworks (\ref{txt:CONST}) and (\ref{txt:IMB}). $P_{k_0}$ is the mid price at the execution time $k_0$. We kept the same initial parameters of Figures \ref{fig:deltaPImbalanceNC_Vs_OC}.a and \ref{fig:deltaPImbalanceNC_Vs_OC}.b.

Figure \ref{fig:deltaPErrorImbalance} deserves the following comments. As expected, the optimal control provides better results than a blind ``join the bid'' strategy.
In Figure \ref{fig:deltaPErrorImbalance}.a the price improvement 
is non-negative since our control maximizes $\Esp_{U_0,\mu} \left( \Delta \text{P}\right)$. When the initial imbalance is highly positive, the price improvement is close to 0 however when the initial imbalance is highly negative the price improvement becomes higher than 0 by avoiding adverse selection. Similarly, Figure \ref{fig:deltaPErrorImbalance}.b shows that the optimal strategy allows to buy with a low average price when imbalance is highly negative by preventing from adverse selection
\footnote{Indirectly, maximizing $\Esp_{U_0,\mu} \left( \Delta \text{P} | \text{Exec} \right)$  leads to the minimization of the price.}.

\subsubsection{Average Duration of Optimal Strategies}

\begin{figure}[!ht]
  \centering
  \hfill (a)\hfill (b) \hfill~\\
  \includegraphics[width=.45\linewidth]{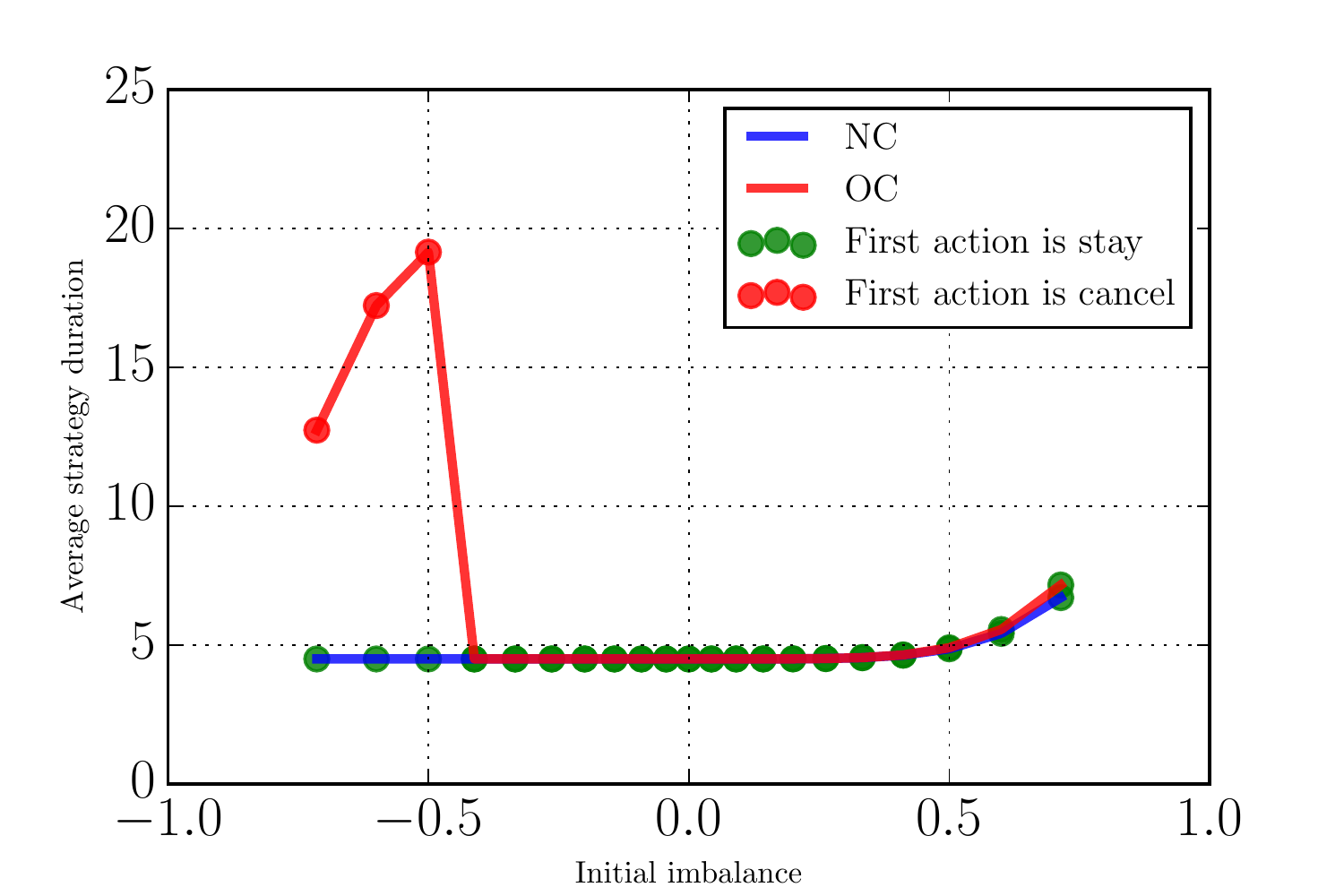}
  \includegraphics[width=.45\linewidth]{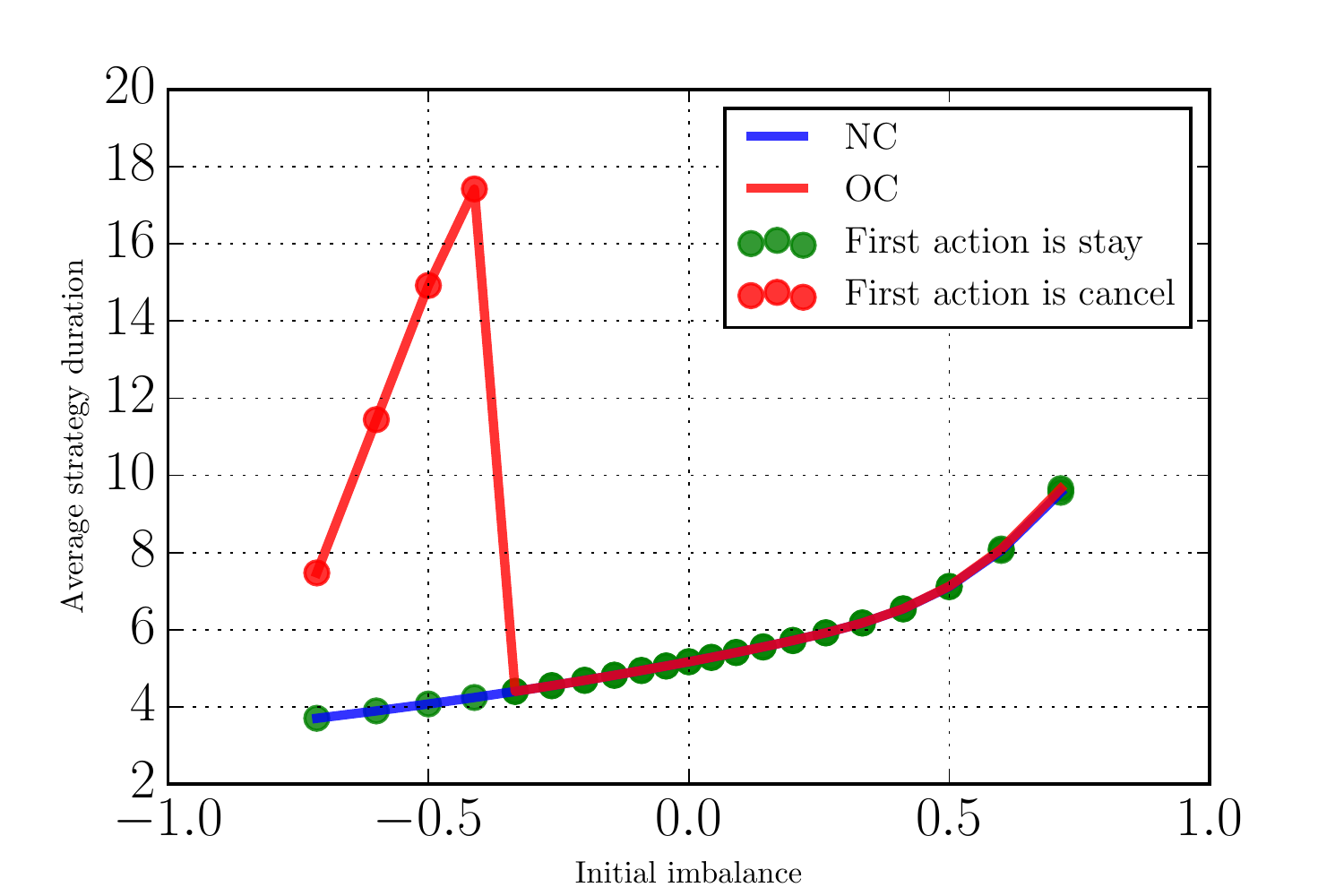}\\
  \hfill (c) \hfill~\\
  \includegraphics[width=.45\linewidth]{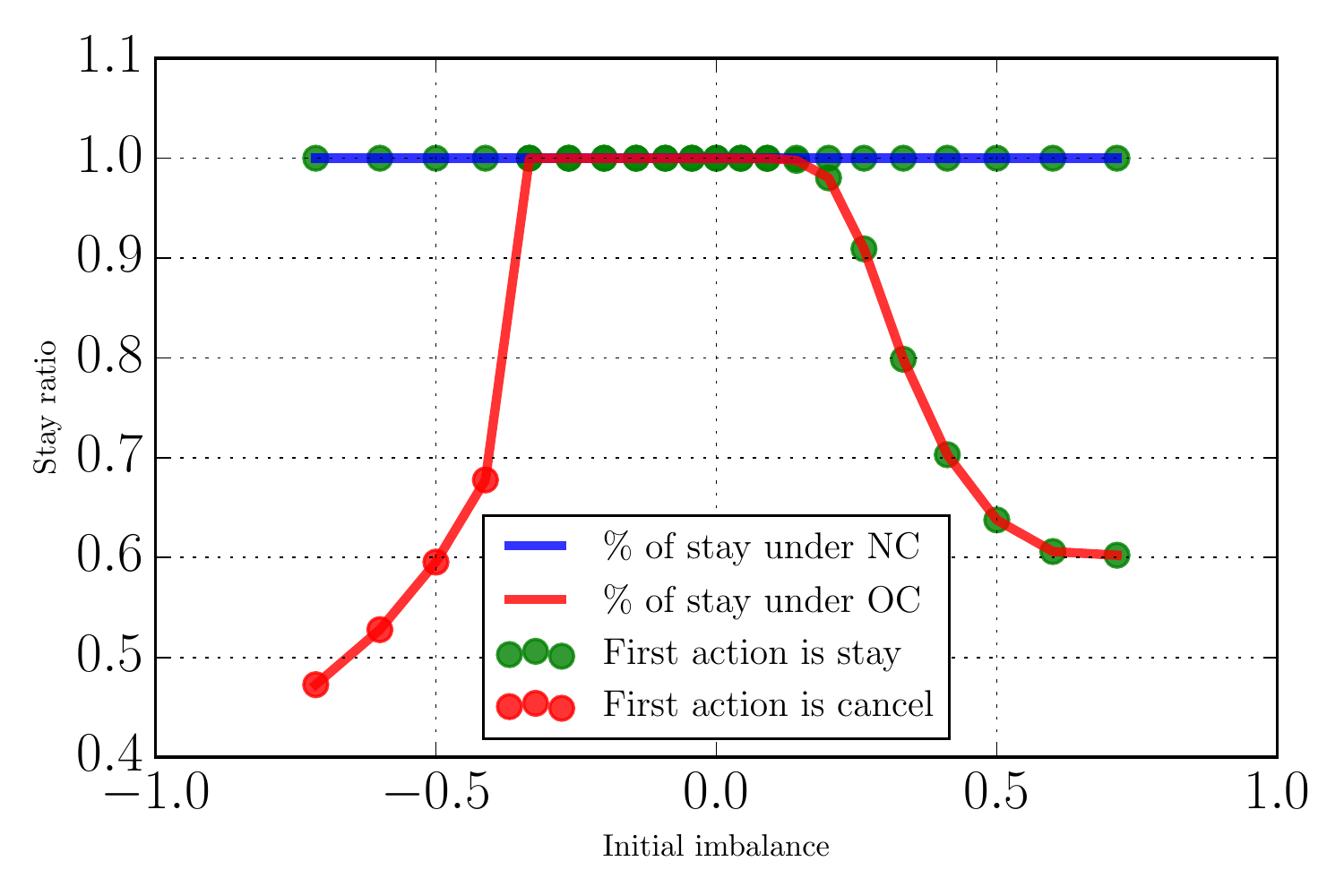}
  \caption{Average strategy duration as a function of (a) the initial imbalance under (\ref{txt:CONST}) and (b) the initial imbalance under (\ref{txt:IMB}). (c) Stay ratio as a function of the initial imbalance under (\ref{txt:IMB}). }
  \label{fig:MeanExecTimeImbalance}
\end{figure}

In brief, the optimal strategy aims to obtain an execution in the best market conditions (i.e. with a low adverse selection risk). It can be read on the average lifetime (i.e. "duration") of the strategy. 
Figure \ref{fig:MeanExecTimeImbalance}.a compares the average strategy duration 
 in both frameworks (\ref{txt:NC}) and (\ref{txt:OC}) when intensities are constant (\ref{txt:CONST}). We kept the same initial parameters of Figure \ref{fig:deltaPImbalanceNC_Vs_OC}.a. Figure \ref{fig:MeanExecTimeImbalance}.b is the analogous of Figure \ref{fig:MeanExecTimeImbalance}.a but under framework (\ref{txt:IMB}). Finally, Figure \ref{fig:MeanExecTimeImbalance}.c shows the "stay ratio" (i.e. the proportion of trajectories for which the optimal strategy chooses to not cancel its limit order) under both frameworks  (\ref{txt:NC}) and (\ref{txt:OC}) when intensities depend on the imbalance (\ref{txt:IMB}). Figure \ref{fig:MeanExecTimeImbalance}.b and \ref{fig:MeanExecTimeImbalance}.c are computed with the same initial parameters of Figure \ref{fig:deltaPImbalanceNC_Vs_OC}.b.

In both Figures \ref{fig:MeanExecTimeImbalance}.a and \ref{fig:MeanExecTimeImbalance}.b, the average strategy duration of the optimal control is always higher than the non-optimal one. It is an expected result since the optimal control cancels the order and hence postpone the execution. Moreover, the algorithm cancels the order when high adverse selection is present (i.e the imbalance is highly negative under \ref{txt:IMB} \footnote{ close to $t=0$, the optimal strategy is free to cancel its limit order; but when $T_f$ is close, it has to think about the cost of having to cross the spread in few steps.}). In such case, the average strategy duration of the optimal control is strictly greater than the non-optimal one (see Figures \ref{fig:MeanExecTimeImbalance}.a and \ref{fig:MeanExecTimeImbalance}.b).

\color{blue}**\color{black}\manuallabel{IMBIntensInf}{**} In Figure \ref{fig:MeanExecTimeImbalance}.b, when intensities depend on the imbalance (\ref{txt:IMB}), the average strategy duration has an increasing trend. In fact, under (\ref{txt:IMB}), when imbalance is highly positive, more weights are given to events delaying the execution. For example, when imbalance is highly positive, the bid queue is a way larger than the opposite one. Then, the probability to obtain an execution on the bid side is low : that's why it is expected to wait more. Moreover, Figure \ref{fig:MeanExecTimeImbalance}.c shows that the agent become more active when high adverse selection is present. Indeed, when the imbalance is negative (i.e. high adverse selection), the "stay ratio" decreases and consequently the "cancel ratio" increases.

\subsubsection{Influence of the Terminal Constraint}

In this section, we want to shed light on two stylized facts:
\begin{enumerate}
\item the optimal strategy performs  better under good market condition when there is more time left.
\item the optimal strategy becomes highly active close to the terminal time.
\end{enumerate}  

\begin{figure}[!ht]
  \centering
  \hfill (a)\hfill (b) \hfill~\\
  \includegraphics[width=.4\linewidth]{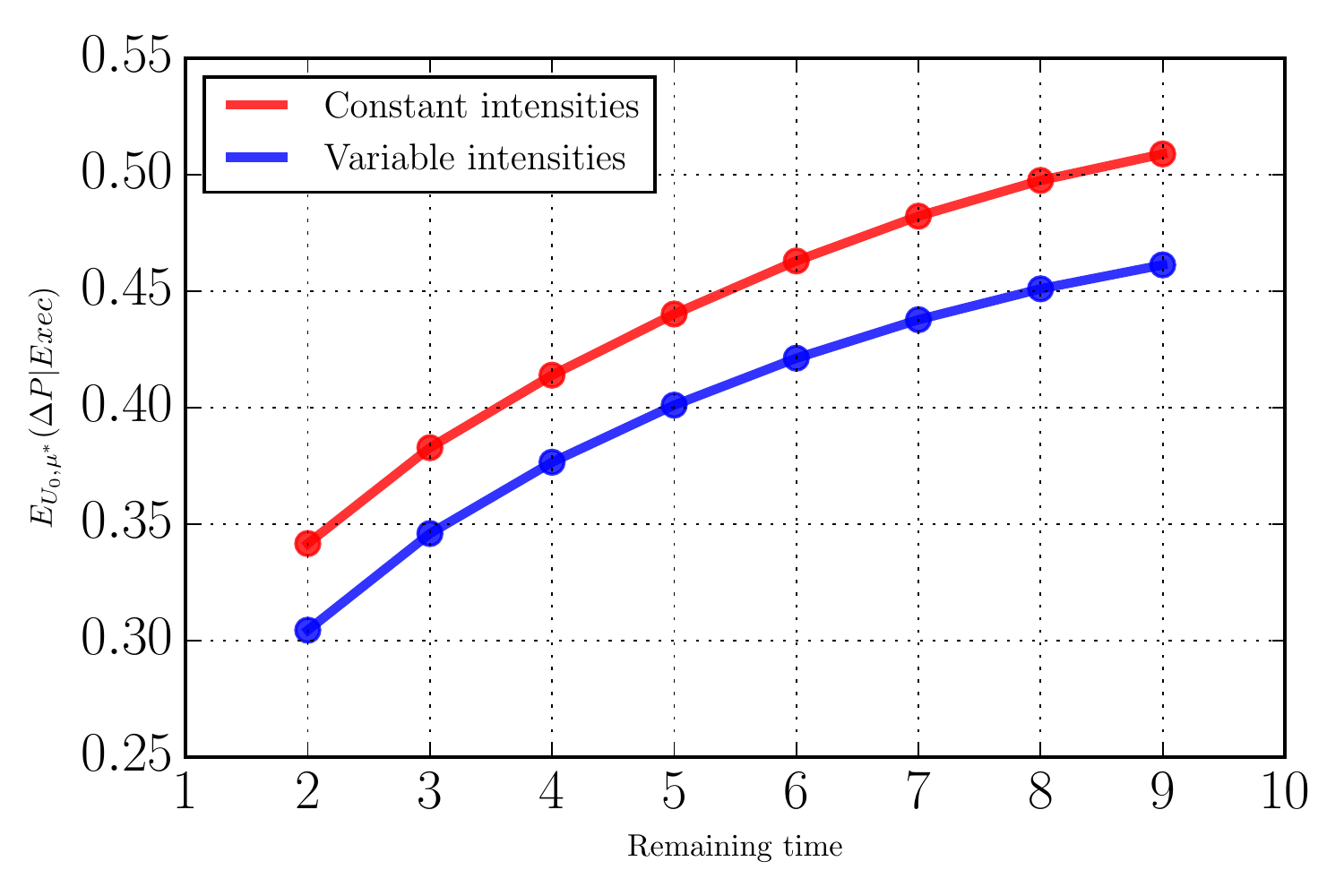}
  \includegraphics[width=.4\linewidth]{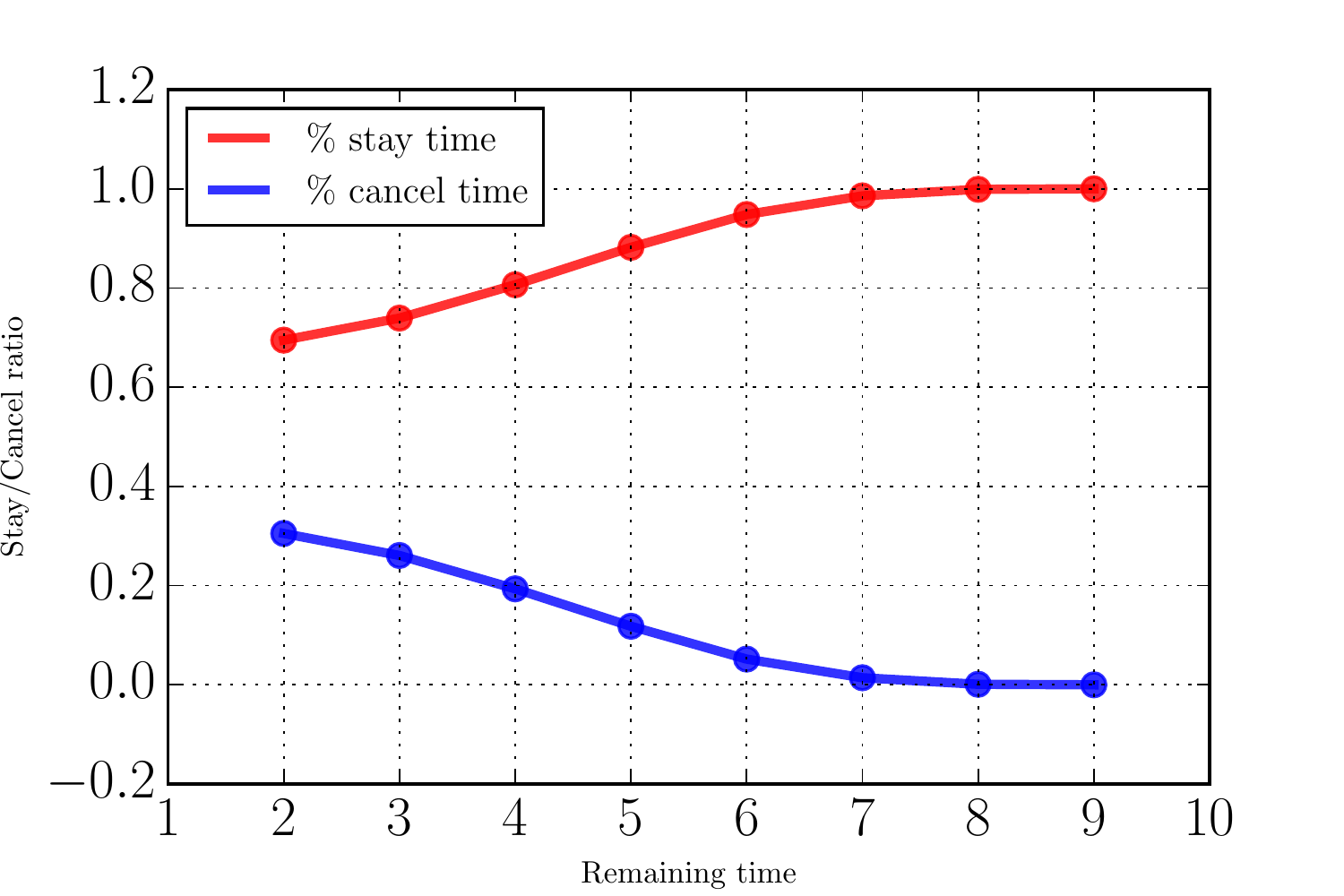}
  \caption{ (a) $\Esp_{U_0,\mu^*} \left( \Delta \text{P} | \text{Exec} \right)$ move relative to remaining time under (\ref{txt:CONST}) and (\ref{txt:IMB}). (b) stay cancel ratio move relative to remaining time to maturity under (\ref{txt:IMB}).}
  \label{fig:deltaPImbalance}
\end{figure}

Figure \ref{fig:deltaPImbalance}.a compares the variation of $\Esp_{U_0,\mu^*} \left( \Delta \text{P} | \text{Exec} \right)$ as a function of the remaining time under frameworks (\ref{txt:CONST}) and (\ref{txt:IMB}). Initial imbalance is fixed equal to 0.5. Thanks to the Figure \ref{fig:deltaPImbalance}.a, we can see that the more time remaining, the better for the optimal strategy. However, the concavity of the curve shows that the marginal performance $ \frac{\partial \Esp \left( \Delta \text{P} \right)_t}{\partial t}  $ is decreasing. Moreover,  Figure \ref{fig:deltaPImbalance}.a shows also that $\Esp_{U_0,\mu^*} \left( \Delta \text{P} | \text{Exec} \right)$ may converge to a limit value when maturity time tends to infinity. Since the markov chain $U_n$ is ergodic  (cf. \cite{citeulike:12810809}), we believe that this limit value is unique and independent of the initial state of the orderbook and could lead to an ``almost ergodic'' regime.

In Figure \ref{fig:deltaPImbalance}.b, we represent the percentage of times where the optimal strategy cancels its order and the percentage of times where it decides to stay in the orderbook as a function of remaining time under (\ref{txt:CONST}) and (\ref{txt:IMB}). The initial imbalance is fixed to 0.5. Thanks to Figure \ref{fig:deltaPImbalance}.b, we conclude that it is optimal to be more active close to $t=T_f$. In Figures \ref{fig:deltaPImbalance}.a and \ref{fig:deltaPImbalance}.b, we kept the same initial parameters of Figures \ref{fig:deltaPImbalanceNC_Vs_OC}.a and \ref{fig:deltaPImbalanceNC_Vs_OC}.b

\subsection{The Price of Latency}
In Section \ref{sec:dpp}, we defined the Markov chain $U^{\mu}_n$ that corresponds to a market participant enabled to change his control at each period. A slower participant will not react at each limit orderbook move. Hence, he can be modelled by the markov chain $U^{\mu}_{\tau n}$ where $\tau$ corresponds to a latency factor such as $\tau \in \mathbb{N}^{*}$.

Using notations of previous sections, we define $Z_{\tau,f}$ as the final constraint associated to the Markov chain $U^{\mu}_{\tau n}$. Thus, we define the latency cost of a participant with a latency factor $\tau$ such as :

\begin{equation}
Latency_{U_0,f} ( \tau)= V_{U_0,f} - V_{U_0,f,\tau} \qquad  \forall\tau \in \mathbb{N}^{*}
\end{equation}
Where $V_{U_0,f,\tau}= \underset{\mu \in \mathcal{U}}{\sup} \Esp_{U_0,\mu} \left(Z_{\tau,f}\right) $.

By adapting the same numerical forward-backward algorithm, the cost of latency can be computed numerically. This cost can be converted into a value: it is the value a trader should accept to pay in technology since he will be rewarded in term of performance. 

\begin{figure}[!ht]
  \centering
  \hfill (a) \hfill (b) \hfill~\\
  \includegraphics[width=.4\linewidth]{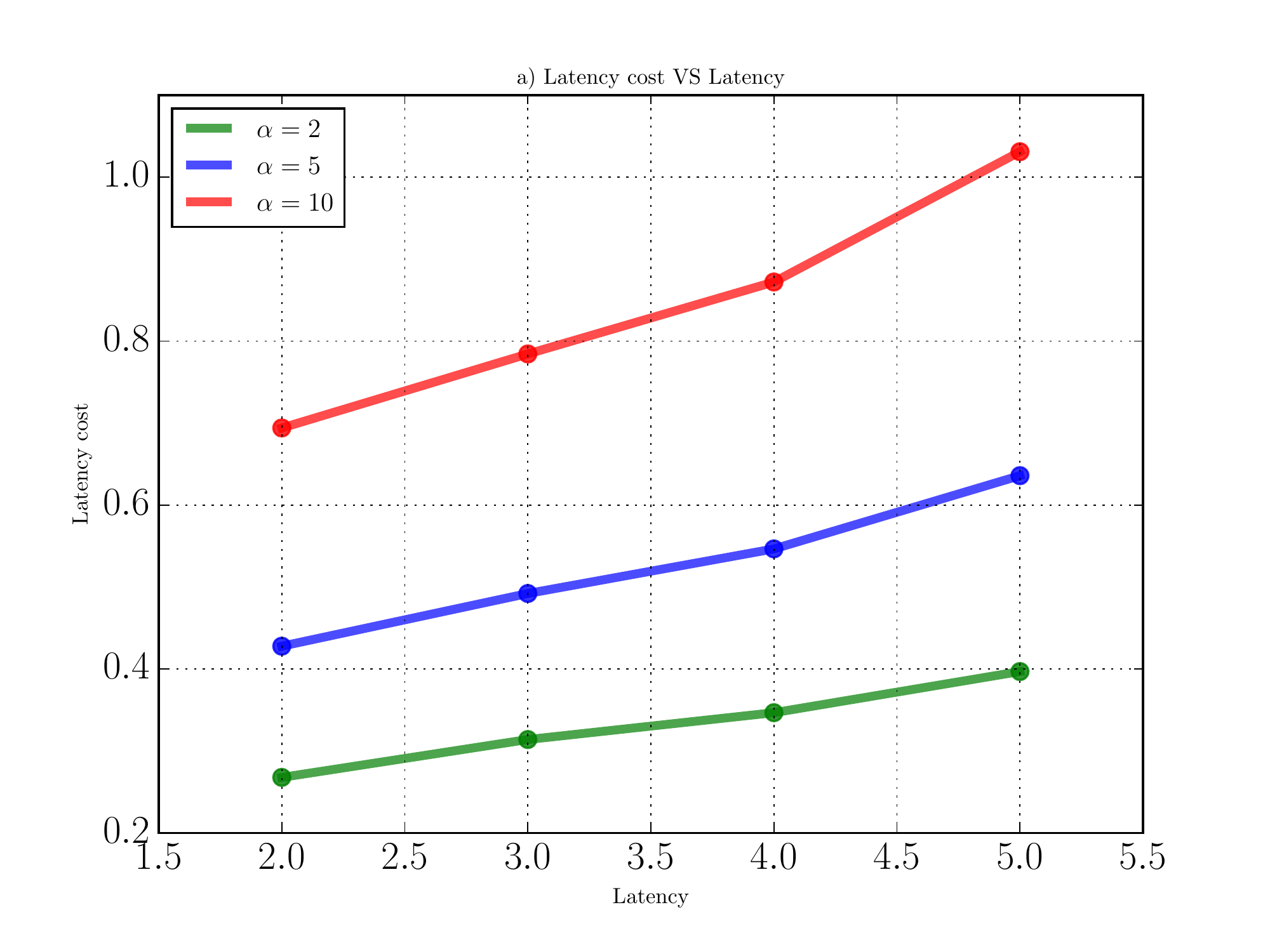}
  \includegraphics[width=.4\linewidth]{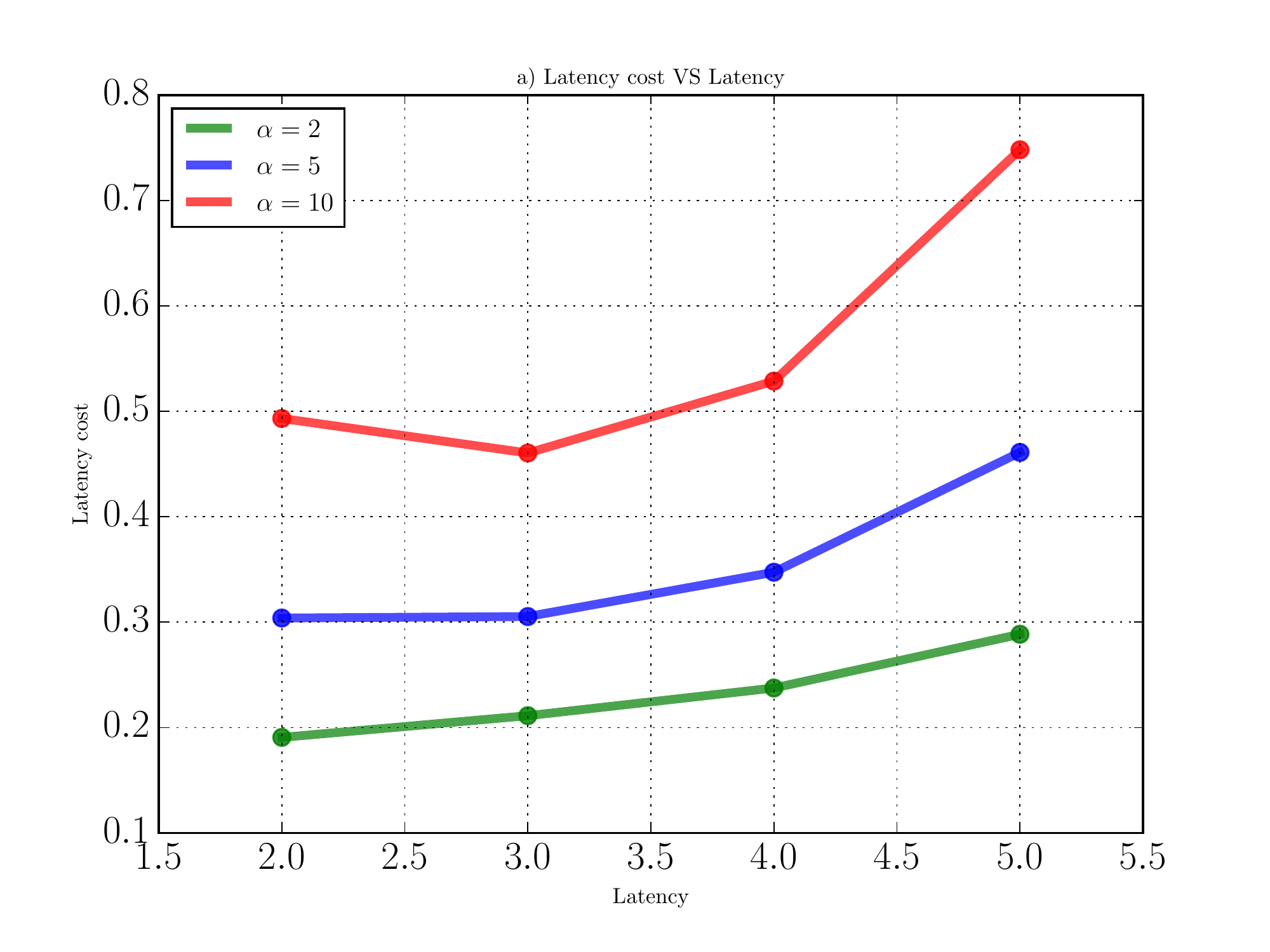}
  \caption{ (a)  Latency cost as a function of the latency factor $\tau$ under (\ref{txt:CONST}). (b) Latency cost as a function of the latency factor $\tau$ under (\ref{txt:IMB}) for different values of $\alpha$.}
    \label{fig:LatencyCost}
\end{figure}

Figures \ref{fig:LatencyCost}.a and \ref{fig:LatencyCost}.b show the variation of the latency cost with respect to the latency factor  $\tau$ under both frameworks  (\ref{txt:CONST}) and (\ref{txt:IMB}) for different values of $\alpha$. The initial imbalance is fixed equal to 0.5 with an initial state $Q^{After}_0=2$,$Q^{Before}_0=1$ and $Q^{Opp}_0=1$. We kept the same initial parameters of Figure \ref{fig:deltaPImbalanceNC_Vs_OC}.a and Figure \ref{fig:deltaPImbalanceNC_Vs_OC}.b.

Numerical results show :
\begin{itemize}
\item The latency cost increases with the latency factor $\tau$ (cf. Figure \ref{fig:LatencyCost}.a). 
\item The latency cost is higher when sensitivity to adverse selection increases (i.e $\alpha$ is big) (cf. Figures \ref{fig:LatencyCost}.a and \ref{fig:LatencyCost}.b).\\
\end{itemize}

Consequently, the added value of exploiting a knowledge on liquidity imbalance is eroded by latency: being able to predict future liquidity consuming flows is of less use if you can't cancel and reinsert your limit orders at each change of the orderbook state. For instance, when two agents act optimally according the same criterion, the faster will have more profits than the slower. 

\bigbreak 
\bigbreak 
\bigbreak 

\section{Conclusion}

We have used NASDAQ-OMX labelled data to show how market participants accept or refuse transactions via limit orders as a function of liquidity imbalance. It is not an exhaustive study on this exchange from the north of Europe (we focus on  AstraZeneca from January 2013 to September 2013). We first show that the orderbook imbalance has a predictive power on future mid price move.
We then focus on three types of market participants: Institutional brokers, Global Investment Banks (GIB) and High Frequency Participants (HFP). Data show that the former accept to trade when the imbalance is more negative (i.e. they buy or sell while the price pressure is downward or upward) than GIB, themselves accepting a more negative imbalance than HFP.
Moreover, when we split HFP between high frequency market makers and high frequency proprietary traders (HFPT), we see that HFPT achieve to buy via limit orders when the imbalance is very small. We complete this analysis with the dynamics of prices around limit orders execution, showing how strategically participants use  their limit orders.

Then we proposed a theoretical framework to control limit orders where liquidity imbalance can be used to predict future price moves. Our framework includes potential adverse selection via a parameter $\alpha$.
We use the dynamic programming principle to provide a way to solve it numerically and exhibit simulations. We show that solutions of our framework have commonalities with our empirical findings.

In a last Section we show how the capability of exploiting imbalance predictability using optimal control decreases with latency: the trader has less time to put in place sophisticated strategies, hence he cannot take profit of any strategy gain. 
\medskip

The difficult point of using limit orders is adverse selection: if the price has chances to go down the probability to be filled is high but it is better to postpone execution to get a better price. However, when a market participant cancels his limit order (to postpone execution), he takes the risk to never obtain a transaction:  the reinsertion of the order will be on the top of the bid queue. Furthermore, the price may go up again before the execution of the limit order.

Our framework includes all these effects and our optimal strategy makes the choice between waiting in the queue or leaving it when the probability the price will go down is too high. To do this, the position of the limit order in the queue is taken into account by our controller.
\smallskip

This leads to a quantitative way to understand market making and latency: if a market maker is fast enough, he will be able to play this insert, cancel and re-insert game to react to his observations of liquidity imbalance. In our framework we use the difference between the sizes of the first bid and ask queues as a proxy of liquidity imbalance, in the real word market participants can use a lot of other information (like liquidity imbalance on correlated instruments, or realtime news feeds).

In such a context speed can be seen as a protection to adverse selection, potentially reducing transaction costs. 
Within this viewpoint, high frequency actions do not add noise to the price formation process (as opposite to the viewpoint of \cite{citeulike:12721030}) but allows market makers to offer better quotes.
At this stage, we do not conclude speed is good for liquidity because:
\begin{itemize}
\item We only focussed on one limit order, we should go towards a framework similar to the one of \cite{citeulike:9304794} to conclude on the added value of imbalance for the whole market making process, but it will be too sophisticated at this stage.
\item It is not fair to draw conclusions from a knowledge of the theoretical optimal behaviour of one market participant; to go further we should model the game played by all participants, similarly to what have been done in \cite{citeulike:12386824}. Again it is a very sophisticated work. Nevertheless, this paper is a first step. We are convinced it is possible to obtain partially explicit formula, to enable more systematic explorations of the influence of parameters (currently our simulations are highly memory consuming). It should allow to confront our results to observed behaviors with accuracy (especially using observed values for our parameters $\alpha,\beta,Q^{Disc},Q^{Ins}$ and $\lambda$s).
\item Last but not least, any conclusion on the added value of low latency and high frequency market making should take into account market conditions. Its value could change with the level of stress of the price formation.
\end{itemize}
\medskip

\noindent This work shows that imbalance is used by participants, and provides a theoretical framework to play with limit order placement. It can be used by practitioners.
More importantly, we hope other researchers will extend our work in different directions to answer to more questions, and we will ourselves continue to work further to understand better liquidity formation at the smallest time scales thanks to this new framework.

\paragraph{Acknowledgements.}
Authors would like to thank Sasha Sto\"\i{}kov and Jean-Philippe Bouchaud for discussions about orderbook dynamics and optimal placement of limit orders that motivated this paper.
Moroever, authors would like to underline the work of Gary Sounigo (during his Masters Thesis) and Felix Patzelt (during a post-doctoral research), who worked hard at Capital Fund Management (CFM) to understand how to align NASDAQ-OMX labeled transactions to direct datafeed of orderbook records. Long lasting discussions about limit order placement with Capital Fund Management execution researchers (especially with Bence Toth and Mihail Vladkov) influenced very positively our work.

\bibliographystyle{apalike}
\bibliography{lehalle}

\newpage

\appendix

\section{Transition probabilities of the markov chain $U_n$}
\label{sec:app:transition}

When first limits are totally consumed, new quantities $Q^{Disc}_n $ and $Q^{Ins}_n$ are inserted in the orderbook. We introduce then $\phi_n $ the joint distribution of the random variables $Q^{Disc}_n $ et $Q^{Ins}_n$ at time $t_n$. We assume these two variables are independent from their past and independent from the counting processes $N^{Same,+}$, $ N^{Same,-}$, $N^{Opp,-}$ and $N^{Opp,+}$. However, $Q^{Disc}_n $ and $Q^{Ins}_n$ can be correlated at time $t_n$. 

Let $ n \in \left\{0 ,1, \cdots ,f \right \} $, $p \in \mathbb{R}_{+}$, $q^{bef} \in \mathbb{N}$, $q^{aft} \in \mathbb{N}$, $q^{opp} \in \mathbb{N}$, 
$q^{disc} \in \mathbb{N}$, $q^{ins} \in \mathbb{N}$ and $e \in \left\{-1, 0, 1 \right\} $

\begin{itemize}
\item \textbf{When the order has been executed before $t_n$ (i.e $e=1$ or $ e=-1$)},then :
\begin{equation*}
\mathbb{P}\left(U_{n+1}=(p,q^{bef},q^{aft},q^{opp},-1) / U_n =(p,q^{bef},q^{aft} ,q^{opp},e) \right) = 1.
\end{equation*}
When the order is executed a dead center is reached and both quantities and the price remain unchanged. In such case, the control has no more infuence.

\item \textbf{When the order isn't executed at $t_n$ (i.e $e=0$)}, and :
\subparagraph{A unit quantity is added to $Q^{Opp}$.} Under control "c", the transition probability is the following : 
\begin{align*}
\mathbb{P}^{(c)}\big(&U_{n+1}=(p,q^{bef},q^{aft},\left(q^{opp}+1\right),e) | U_n=(p,q^{bef},q^{aft} ,q^{opp},e) \big) \\
& = \mathbb{P}  ( \left\{N^{Opp,+}_{n+1}-N^{Opp,+}_{n} =1 \right\} \cap \left\{N^{Opp,-}_{n+1} - N^{Opp,-}_{n} =0 \right\} \\
&\quad \quad \cap \left\{N^{Same,+}_{n+1}-N^{Same,+}_{n} =0\right\}  \cap \left\{N^{Same,-}_{n+1}-N^{Same,-}_{n} =0\right\}) \\ 
 & =  \lambda^{Opp,+}_n\Delta_t \left(1-\lambda^{Opp,-}_n \Delta_t \right)\left (1-\lambda^{Same,-}_n \Delta_t \right) \left (1-\lambda^{Same,+}_n \Delta_t \right).
\end{align*}
Under control "s" :
\begin{align*}
\mathbb{P}^{(s)}\big(&U_{n+1}=(p,\left(q^{bef}+q^{aft}\right),0,\left(q^{opp}+1\right),e) | U_n=(p,q^{bef},q^{aft} ,q^{opp},e) \big) \\
 & =  \lambda^{Opp,+}_n\Delta_t \left(1-\lambda^{Opp,-}_n \Delta_t \right)\left (1-\lambda^{Same,-}_n \Delta_t \right) \left (1-\lambda^{Same,+}_n \Delta_t \right).
\end{align*}

\subparagraph{A unit quantity is cancelled from $Q^{Opp}$.} We differentiate between two cases: 
\begin{enumerate}
\item \textbf{When $q^{opp}>1$},  under control "c" :
	\begin{align*}
\mathbb{P}^{(c)}\big(&U_{n+1}=(p,q^{bef},q^{aft},\left(q^{opp}-1\right),e) | U_n=(p,q^{bef},q^{aft} ,q^{opp},e) \big)  \\
& = \mathbb{P}  ( \left\{N^{Opp,+}_{n+1}-N^{Opp,+}_{n} =0 \right\} \cap \left\{N^{Opp,-}_{n+1}-N^{Opp,-}_{n} =1 \right\} \\
& \quad \quad \cap \left\{N^{Same,+}_{n+1}-N^{Same,+}_{n} =0\right\}  \cap \left\{N^{Same,-}_{n+1}-N^{Same,-}_{n} =0\right\}) \\
& = \lambda^{Opp,-}_n\Delta_t \left(1-\lambda^{Opp,+}_n \Delta_t \right)\left (1-\lambda^{Same,-}_n \Delta_t \right) \left(1-\lambda^{Same,+}_n \Delta_t \right).
    \end{align*}
Under control "s":
	\begin{align*}
\mathbb{P}^{(s)}\big(&U_{n+1}=(p,\big(q^{bef}+q^{aft}\big),0,\left(q^{opp}-1\right),e) | U_n=(p,q^{bef},q^{aft} ,q^{opp},e) \big)  \\
& = \lambda^{Opp,-}_n\Delta_t \left(1-\lambda^{Opp,+}_n \Delta_t \right)\left (1-\lambda^{Same,-}_n \Delta_t \right) \left(1-\lambda^{Same,+}_n \Delta_t \right).
    \end{align*}
\item  \textbf{When $q^{opp}\leq1$}, the price increases by one tick under control 'c' :
  \begin{equation*}
    \begin{split}
      \mathbb{P}^{(c)}\big(&U_{n+1} = (p+1,q^{ins},0,q^{disc},e)\; |\; U_n=(p,q^{bef},q^{aft} ,1,e) \big) \\ &=\lambda^{Opp,-}_n\Delta_{t} \left(1-\lambda^{Opp,+}_n \Delta_t \right)\left (1-\lambda^{Same,-}_n \Delta_t \right) \left (1-\lambda^{Same,+}_n \Delta_t \right) \phi_{n+1} (q^{disc},q^{ins}) .
    \end{split}
  \end{equation*}
Under control "s", the last formula does not change.
\end{enumerate}
\subparagraph{A unit quantity is added to $Q^{Same}$}, under control "c" we have : 
\begin{align*}
\mathbb{P}^{(c)}\big(&U_{n+1}=(p,q^{bef},q^{aft}+1,q^{opp},e) | U_n=(p,q^{bef},q^{aft},q^{opp},e) \big) \\
&= \lambda^{Same,+}_n\Delta_t \left(1-\lambda^{Opp,+}_n \Delta_t \right)\left (1-\lambda^{Opp,-}_n \Delta_t \right) \left (1-\lambda^{Same,-}_n \Delta_t \right).
\end{align*} 
Under control "s": 
\begin{align*}
\mathbb{P}^{(s)}\big(&U_{n+1}=(p,\big(q^{bef}+q^{aft}+1\big),0,q^{opp},e) | U_n=(p,q^{bef},q^{aft},q^{opp},e) \big) \\
&= \lambda^{Same,+}_n\Delta_t \left(1-\lambda^{Opp,+}_n \Delta_t \right)\left (1-\lambda^{Opp,-}_n \Delta_t \right) \left (1-\lambda^{Same,-}_n \Delta_t \right).
\end{align*} 

\subparagraph{A unit quantity is cancelled from $Q^{Same}$.} We distinguish again three cases : 
\begin{enumerate}
\item \textbf{When $\left( q^{bef}>1 \, \text{and} \,  q^{aft}\geq 0 \right)$ or $\left( q^{bef}=1 \, \text{and} \,  q^{aft}\geq 1 \right)$}, under control "c":
\begin{align*}
\mathbb{P}^{(c)}\big(&U_{n+1}=(p,q^{bef}-1,q^{aft},q^{opp},e) | U_n=(p,q^{bef},q^{aft},q^{opp},e) \big) \\
& = \lambda^{Same,-}_n\Delta_t\left(1-\lambda^{Opp,+}_n \Delta_t \right)\left (1-\lambda^{Opp,-}_n \Delta_t \right) \left (1-\lambda^{Same,+}_n \Delta_t \right).
\end{align*}
We suppose that our order is executed when $Q^{After}_n$ is consumed. Under control "s":
\begin{align*}
\mathbb{P}^{(s)}\big(&U_{n+1}=(p,\big(q^{bef}+q^{aft}-1\big),0,q^{opp},e) | U_n=(p,q^{bef},q^{aft},q^{opp},e) \big) \\
& = \lambda^{Same,-}_n\Delta_t\left(1-\lambda^{Opp,+}_n \Delta_t \right)\left (1-\lambda^{Opp,-}_n \Delta_t \right) \left (1-\lambda^{Same,+}_n \Delta_t \right).
\end{align*}
\item \textbf{When $q^{bef}=0 \, \text{and} \, q^{aft}> 1$}, under control "c":
\begin{align*}
\mathbb{P}^{(c)}\big(&U_{n+1}=(p,0,\left(q^{aft}-1\right),q^{opp},1) | U_n=(p,q^{bef},q^{aft},q^{opp},e) \big) \\
& = \lambda^{Same,-}_n\Delta_t \left(1-\lambda^{Opp,+}_n \Delta_t \right)\left (1-\lambda^{Opp,-}_n \Delta_t \right) \left (1-\lambda^{Same,+}_n \Delta_t \right).
\end{align*}
Under control "s":
\begin{align*}
\mathbb{P}^{(s)}\big(&U_{n+1}=(p,\left(q^{aft}-1\right),0,q^{opp},0) | U_n=(p,q^{bef},q^{aft},q^{opp},e) \big) \\
& = \lambda^{Same,-}_n\Delta_t \left(1-\lambda^{Opp,+}_n \Delta_t \right)\left (1-\lambda^{Opp,-}_n \Delta_t \right) \left (1-\lambda^{Same,+}_n \Delta_t \right).
\end{align*} 
\item \textbf{When $ q^{bef} + q^{aft} = 1 $}, under control "c" :
\begin{align*}
\mathbb{P}^{(c)}\big(&U_{n+1}=(p,q^{disc},0,q^{ins},1) | U_n=(p,q^{bef},q^{aft},q^{opp},e) \big) \\
& = \lambda^{Same,-}_n\Delta_t \left(1-\lambda^{Opp,+}_n \Delta_t \right)\left (1-\lambda^{Opp,-}_n \Delta_t \right) \left (1-\lambda^{Same,+}_n \Delta_t \right) \phi_{n+1} (q^{disc},q^{ins}).
\end{align*}
Under control "s":  
\begin{align*}
\mathbb{P}^{(s)}\big(&U_{n+1}=(p,q^{disc},0,q^{ins},0) | U_n=(p,q^{bef},q^{aft},q^{opp},e) \big) \\
& = \lambda^{Same,-}_n\Delta_t \left(1-\lambda^{Opp,+}_n \Delta_t \right)\left (1-\lambda^{Opp,-}_n \Delta_t \right) \left (1-\lambda^{Same,+}_n \Delta_t \right) \phi_{n+1} (q^{disc},q^{ins}).
\end{align*}
\end{enumerate}

\subparagraph{Nothing happens } in the limit order book with probability 
\begin{equation*}
\mathbb{P}^{(c)}\big(U_{n+1}=U_n| U_n \big)  = \left(1-\lambda^{Same,-}_n\Delta_t \right) \left(1-\lambda^{Opp,+}_n \Delta_t \right)\left (1-\lambda^{Opp,-}_n \Delta_t \right) \left (1-\lambda^{Same,+}_n \Delta_t \right).
\end{equation*}

\item For all the remaining cases we assume the transition probability neglectibe. We hence set it to zero.
\end{itemize}

\subparagraph{Remark.} 
By taking in consideration the different cases and neglecting the terms with order strictly superior than 1 in $\Delta_t$, we have for any control $ i \in \{c,s\}$: 
\begin{align*}
\sum_{\begin{array}{c}\text{states} \, U_n\\\text{states} \, U_{n+1}\end{array}} \int_{(\mathbb{N}_{+})^2} \mathbb{P}^{(i)}(U_{n+1}| U_n) \;\mu_{n+1} (\mathrm{d}\, q^{disc},\mathrm{d} \, q^{ins}) \\
\approx 1 + \lambda^{Same,+}_n\Delta_t + \lambda^{Same,-}_n\Delta_t +\lambda^{Opp,+}_n\Delta_t +\lambda^{Opp,-}_n\Delta_t .
\end{align*}
Consequently, if $ \lambda^{Same,+}_n\Delta_t + \lambda^{Same,-}_n\Delta_t +\lambda^{Opp,+}_n\Delta_t +\lambda^{Opp,-}_n\Delta_t = \text{o} (1)$ ( which is true when $\Delta_t $ is small), we end up  with for any control $ i \in \{c,s\}$: 
\begin{equation*}
\sum_{\begin{array}{c}\text{states} \, U_n\\\text{states} \, U_{n+1}\end{array}}  \int_{{(\mathbb{N}_{+})^2}} \mathbb{P}^{(i)}(U_{n+1}/U_n) \phi_{n+1} (\mathrm{d} \, q^{disc},\mathrm{d} \, q^{ins}) \approx 1.
\end{equation*} 

\clearpage
\section{Composition of market participants groups}

\begin{table}[!ht]
  \centering
  {\bf High Fequency Traders}\\\medskip
  \begin{tabular}{|rc|cc|}\hline
Name & NASADQ-OMX & Market & Prop. \\
& member code(s)& Maker & Trader\\\hline
    All Options International B.V. & AOI& &\\
    Hardcastle Trading AG & HCT& &\\
    IMC Trading B.V & IMC, IMA& Yes & \\
    KCG Europe Limited & KEM, GEL& Yes & \\
    MMX Trading B.V & MMX& & \\
    Nyenburgh Holding B.V. & NYE& &\\
    Optiver VOF & OPV& & Yes\\
    Spire Europe Limited & SRE, SREA, SREB& & Yes\\
    SSW-Trading GmbH & IAT& & \\
    WEBB Traders B.V & WEB& &\\
    Wolverine Trading UK Ltd & WLV& &\\ \hline
  \end{tabular}
  \caption{Composition of the group of HFT used for empirical examples, and the composition of our ``high frequency market maker'' and ``high frequency proprietary traders'' subgroups.}
  \label{tab:compagents:HFT}
\end{table}

\begin{table}[!ht]
  \centering
  {\bf Global Investment Banks}\\ \medskip
  \begin{tabular}{|rc|}\hline
Name & NASADQ-OMX \\
& member code(s)\\\hline
 Barclays Capital Securities Limited Plc & BRC \\
    Citigroup Global Markets Limited & SAB \\
    Commerzbank AG & CBK \\
    Deutsche Bank AG & DBL \\
    HSBC Bank plc & HBC \\
    Merrill Lynch International & MLI \\
    Nomura International plc & NIP \\\hline
  \end{tabular}
  \caption{Composition of the group of Global Investment Banks used for empirical examples.}
  \label{tab:compagents:GIB}
\end{table}

\begin{table}[!ht]
  \centering
  {\bf Institutional Brokers}\\ \medskip
  \begin{tabular}{|rc|}\hline
Name & NASADQ-OMX\\
& member code(s)\\\hline
 ABG Sundal Collier ASA & ABC \\
              Citadel Securities (Europe) Limited & CDG \\
              Erik Penser Bankaktiebolag & EPB \\
              Jefferies International Limited & JEF \\
              Neonet Securities AB & NEO \\
              Remium Nordic AB & REM \\
              Timber Hill Europe AG & TMB \\\hline
  \end{tabular}
  \caption{Composition of the group of Institutional Brokers used for empirical examples.}
  \label{tab:compagents:brok}
\end{table}

\section{Extreme Imbalances}
\label{sec:extremb}

\begin{figure}[!h]
\begin{minipage}[c]{0.55\linewidth}
\begin{tikzpicture}[scale=0.50]
\draw (2,0) rectangle (3,-5);
\draw [black,fill=gray!60] (2,-4.7) rectangle (3,-5);
\draw (7,-4.7) rectangle (8,-5);
\draw [dotted][->](0,-5) -- ++(11,0);
\draw (5,-5) node{$|$};
\draw[red,fill=red] (6.75,-5) circle (0.25);
\draw [solid][->](6.75,-6.5) -- ++(0,0.75);
\draw [solid][->](2.5,-6.5) -- ++(0,0.75);
\draw [solid][<->](2.75,-6.5) -- ++(3.75,0);
\footnotesize
\draw (2,-4.7) node[left]{$ \text{q} $} ;
\draw (2,-2) node[left]{$Q^{After}$} ;
\draw (8,-4.7) node[right]{$Q^{Same}$} ;
\draw (2.5,-5) node[below]{$ Same $} ;
\draw (7.5,-5) node[below]{$ Opp $} ;
\draw (11,-5) node[below]{$Price$} ;
\tiny
\draw (5,-5) node[below]{$ P(t) $} ;
\draw (2.5,-6.5) node[below]{$P_{\text{Exec}}$} ;
\draw (6.75,-6.5) node[below]{$P_{\infty}$} ;
\draw (4.75,-6.5) node[below]{$\Delta \text{P} >0 $} ;
\normalsize
\draw (3,1) node[right]{if order executed before $T_f$ (1)} ;
\end{tikzpicture}\\

\end{minipage} \hfill
\begin{minipage}[c]{.45\linewidth}
\begin{tikzpicture}[scale=0.45]
\draw [dotted] (1,0) rectangle (2,-5);
\draw [black,fill=gray!60] (1,-4.7) rectangle (2,-5);
\draw [dotted](5,-4.7) rectangle (6,-5);
\draw (7,-3) rectangle (8,-5);
\draw (11,-2) rectangle (12,-5);
\draw [double][->](3.5,-2) -- ++(0.5,0);
\draw [dotted][->](0,-5) -- ++(14,0);
\draw (3.5,-5) node[sloped]{$|$};
\draw (9.5,-5) node[sloped]{$|$};
\draw[red,fill=red] (9.3,-5) circle (0.2);
\draw [solid][->](7.5,-6.5) -- ++(0,0.75);
\draw [solid][->](9.3,-6.5) -- ++(0,0.75);
\draw [solid][<->](7.6,-6.5) -- ++(1.6,0);
\tiny
\draw (3.5,-5) node[below]{$P(t)$} ;
\draw (7,-6.5) node[below]{$P_{\text{Exec}}$} ;
\draw (9.8,-6.5) node[below]{$P_{\infty}$} ;
\draw (8.6,-7.1) node[below]{$\Delta \text{P}>0$};
\footnotesize
\draw (1.5,-5) node[below]{$ Same $} ;
\draw (5.5,-5) node[below]{$ Opp $} ;
\draw (11.5,-2) node[above]{$Q^{Disc}$} ;
\draw (7.5,-3) node[above]{$Q^{Ins}$} ;
\draw (5.5,-4.7) node[above]{$Q^{Opp}$} ;
\draw (1,-2) node[left]{$Q^{After}$} ;
\draw (1,-4.7) node[left]{$ \text{q} $} ;
\draw (13.5,-5) node[below]{$Price$} ;
\normalsize
\draw (3,1) node[right]{if order executed at $T_f$ (2)} ;
\end{tikzpicture}\\
\end{minipage}
\caption{$\Delta \text{P}$ when Imbalance highly positive}
\label{fig:deltaPImbalanceSignPos}
\end{figure}
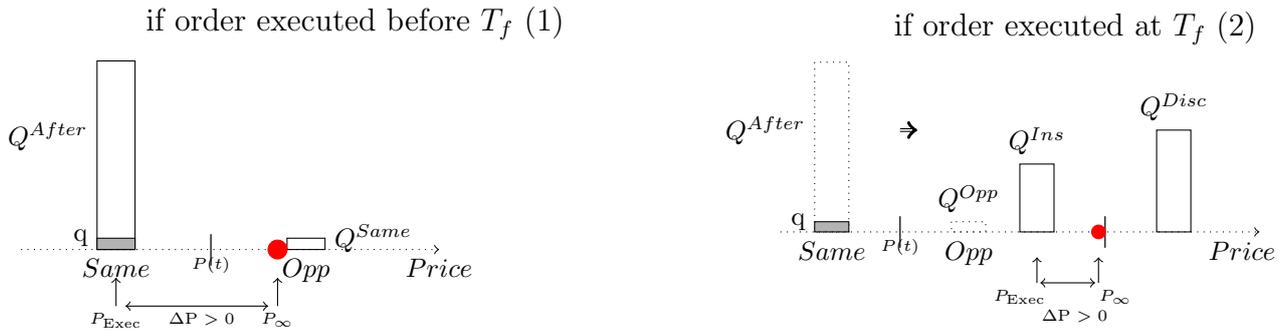

\textbf{The decreasing slope at the right of the curve} in Figure \ref{fig:deltaPImbalanceNC_Vs_OC}.a and \ref{fig:deltaPImbalanceNC_Vs_OC}.b when imbalance is highly positive (i.e $Q^{Same} \gg Q^{Opp} \approx 1$ ). In this situation, the order will be executed in general before the final time $T_f$ without being followed by a price move (1) or will be executed at $T_f$ and followed by a price move (2). In both cases, the final constraint  $\Delta \text{P} $ is positive (see graph \ref{fig:deltaPImbalanceSignPos}). Given that $\Delta \text{P}$ in case (2) is lower than $\Delta \text{P}$ in case (1) and situation (2) occurs more frequently when imbalance is highly positive, it is expected to find a decreasing slope at the right of the curve.

The linear increasing trend of the curve can also be explained by the expression of $\Delta P = \pm 0.5 + \frac{\alpha}{2} \times Imb $ especially when the imbalance effect is not significant (far from the extreme points).  The linear increasing trend of the curve is coherent with the  empirical result in Figure \ref{fig:predpowimb}.

\end{document}